\documentclass[preprint,12pt]{elsarticle}

\usepackage{graphicx}
\usepackage{longtable}

 \usepackage{epsfig}

\usepackage{amssymb}

\usepackage{multirow}
\usepackage[flushleft]{threeparttable}

\usepackage{lineno}
\usepackage{lineno}
\usepackage{multirow}
\usepackage{booktabs}
\usepackage{amsbsy}
\usepackage{soul}
\usepackage{color}

\journal{ApJ}

\begin{document}

\begin{frontmatter}


\title{Search for time-independent neutrino emission from astrophysical sources with 3 years of IceCube data.}

\author[Adelaide]{M.~G.~Aartsen}
\author[MadisonPAC]{R.~Abbasi}
\author[Gent]{Y.~Abdou}
\author[Zeuthen]{M.~Ackermann}
\author[Christchurch]{J.~Adams}
\author[Geneva]{J.~A.~Aguilar}
\author[MadisonPAC]{M.~Ahlers}
\author[Berlin]{D.~Altmann}
\author[MadisonPAC]{J.~Auffenberg}
\author[Bartol]{X.~Bai\fnref{SouthDakota}}
% ...
\address[Adelaide]{School of Chemistry \& Physics, University of Adelaide, Adelaide SA, 5005 Australia}
% ...
\address[Berlin]{Institut f\"ur Physik, Humboldt-Universit\"at zu Berlin, D-12489 Berlin, Germany}
% ...
\address[Christchurch]{Dept.~of Physics and Astronomy, University of Canterbury, Private Bag 4800, Christchurch, New Zealand}
% ...
\address[Geneva]{D\'epartement de physique nucl\'eaire et corpusculaire, Universit\'e de Gen\`eve, CH-1211 Gen\`eve, Switzerland}
\address[Gent]{Dept.~of Physics and Astronomy, University of Gent, B-9000 Gent, Belgium}
% ...
\address[MadisonPAC]{Dept.~of Physics and Wisconsin IceCube Particle Astrophysics Center, University of Wisconsin, Madison, WI 53706, USA}
% ...
\address[Bartol]{Bartol Research Institute and Department of Physics and Astronomy, University of Delaware, Newark, DE 19716, USA}
% ...
\address[Zeuthen]{DESY, D-15735 Zeuthen, Germany}
\fntext[SouthDakota]{Physics Department, South Dakota School of Mines and Technology, Rapid City, SD 57701, USA}
\fntext[LosAlamos]{Los Alamos National Laboratory, Los Alamos, NM 87545, USA}
\fntext[Bari]{also Sezione INFN, Dipartimento di Fisica, I-70126, Bari, Italy}
\fntext[Goddard]{NASA Goddard Space Flight Center, Greenbelt, MD 20771, USA}

\begin{abstract}
  We present the results of a search for neutrino point sources using the
  IceCube data collected between April 2008 and May 2011 with three partially completed
  configurations of the detector: the 40-, 59- and 79-string configurations. The live-time of this data set are 1,040 days.
  An unbinned maximum likelihood ratio test was used to search for an excess of neutrinos
  above the atmospheric background at any given direction in the sky. By adding
  two more years of data with improved event selection and reconstruction techniques,
  the sensitivity was improved by a factor 3.5 or more with respect to the previously published results \cite{IC40}
  obtained with the 40-string configuration of IceCube. We
  performed an all-sky survey and a dedicated search using a catalog of
  \textit{a priori} selected objects observed by other telescopes. In both searches, the
  data are compatible with the background-only hypothesis. In the absence of
  evidence for a signal, we set upper limits on the flux of muon
  neutrinos. For an E$^{-2}$ neutrino spectrum, the observed limits are between $0.9$ and $23.2\times
  10^{-12}$ TeV$^{-1}$ cm$^{-2}$s$^{-1}$. We also report upper limits for neutrino emission from
  groups of sources which were selected according to theoretical models or
  observational parameters and analyzed with a stacking approach.

\end{abstract}


\end{frontmatter}

\section{Introduction}
\label{sec1}

The origin and the acceleration mechanisms of Cosmic Rays (CRs) are a yet
unresolved puzzle. The random-walk of CR particles through
the intergalactic magnetic fields makes it difficult to identify the
cosmological sources of CRs except at the highest energies.
On the other hand, neutrinos are likely produced
in the same environments as CRs and gamma-rays and, being
electrically neutral, they propagate directly from the source to the
Earth. Since these astrophysical neutrinos also have directional information,
their detection will make it possible to unequivocally identify the sources
of CRs. Possible sources may be Supernova Remnant (SNR) shocks,
Active Galactic Nuclei (AGN) jets or Gamma-Ray Bursts (GRBs)~\cite{rev1, rev2}.
The detection of high-energy cosmic
neutrinos will provide a direct proof of hadronic particle acceleration in
the Universe since they can only be produced by the interactions of protons
or nuclei with ambient radiation or matter.

In this paper we present the latest results of the search for neutrino point
sources with the IceCube neutrino observatory. The analysis was done on the
data collected from 2008 to 2011. This paper concerns the searches for steady
neutrino sources while optimized searches for time-dependent neutrino
emission are reported elsewhere~\cite{asen}.

Section~\ref{sec2} describes the IceCube detector and the detection principle.
The three data samples and the corresponding event selections are discussed in Sec.~\ref{sec3}. The
methodology used to combine data from different years and detector
configurations in a point source search is given in Sec.~\ref{sec4} while
Sec.~\ref{sec5} presents the results of the analysis, including a discussion
of their impact on some recent astrophysical models of neutrino
emission. The systematic uncertainties are described in Sec.~\ref{sec6} and conclusions are drawn in Sec.~\ref{sec7}.

\section{Detector}
\label{sec2}

The IceCube detector at the South Pole is designed to observe neutrinos of astrophysical origin and atmospheric muons
and neutrinos induced by Cosmic Rays at the energies around and above the
{\it knee} ($\sim 3 \times 10^{15}$ eV). 

IceCube detects the Cherenkov light emitted by secondary leptons
which are produced in charged current neutrino interactions with the matter surrounding the
detector and is hence sensitive to all neutrino flavors. 
For neutrino point source searches we select events from charged current interactions of muon neutrinos since they result in secondary muons with long tracks and a good directional reconstruction. Above TeV energies,
the scattering angle between the muon and the incoming neutrino is smaller than the angular resolution of the detector. 
In order to detect the Cherenkov light, IceCube
uses an array of 5,160 Digital Optical Modules (DOMs) ~\cite{PMT} deployed on
86 strings at a depth of 1.5-2.5 km below the surface just above the bedrock
in the clear, deep ice. The DOMs are spherical, pressure resistant glass
housings containing each a 25 cm diameter Hamamatsu photomultiplier tube
(PMTs) and electronics for waveform digitization \cite{daq}. High quantum
efficiency PMTs are used in a denser sub-array located in the center of the
detector. This sub-array, called DeepCore, enhances the sensitivity to low
energy neutrinos \cite{DeepCore}. A surface cosmic-ray detector, called
IceTop, completes the IceCube Observatory \cite{icetop}. It uses 324 PMTs to
detect the electromagnetic component of air-showers produced by cosmic-ray
interactions in the atmosphere.

\begin{figure}[!t]
  \vspace{5mm}
  \centering
  \includegraphics[width=4.in]{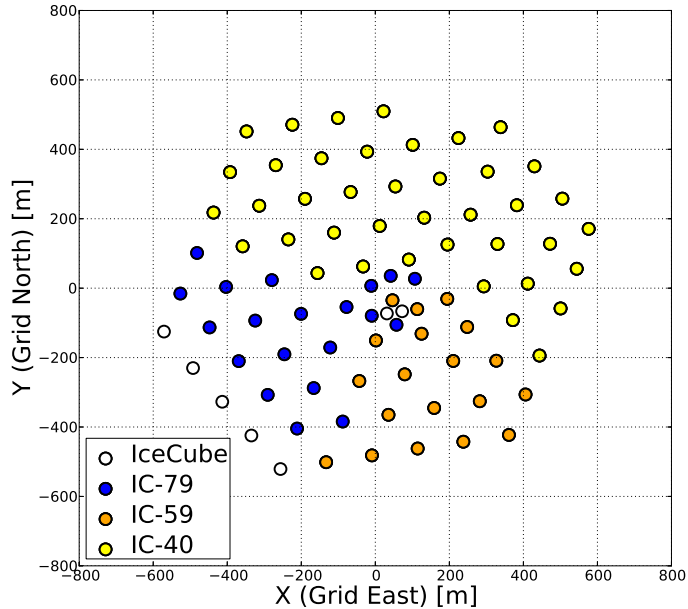}
  \caption{Detector layout in the IceCube coordinate system: The Grid North
    axis is aligned with the Prime Meridian, pointing towards Greenwich,
    UK. The Grid East axis points 90 degrees clock-wise from Grid North. The
    circles represent the surface string positions corresponding to the
    final geometry of the whole IceCube detector. The IC-40 configuration is
    represented by yellow dots. The green circles represent the additional
    strings that form the IC-59 configuration. The IC-59 configuration
    together with the strings indicated by blue circles represent the IC-79
    configuration. The empty circles are the strings added for the complete
    detector.}
  \label{fig:layout}
 \end{figure}

The construction of the IceCube Observatory started in the Austral summer of
2004 and ended in December 2010. Data acquisition with the complete
configuration started in May 2011. However, IceCube has been providing
physics results since the completion of the 9-string array in 2006. From
April 2008 to May 2011 three different configurations of the IceCube detector
were in operation. Fig.~\ref{fig:layout} shows the positions of the strings in
the 40-string configuration (IC-40) which took data from 2008 April 5 to 2009 May 20,
the 59-string configuration (IC-59) active from 2009 May 20 to 2010 May 31,
and the 79-string configuration (IC-79) active
from 2010 May 31 to 2011 May 13. Also shown is the final 86-string IceCube
configuration. The total live-time over the entire period used in this work
corresponds to 1,040 days collected with the IC-40, IC-59 and IC-79
configurations and the average up-time is of $92$\% at final analysis
level.

 In this analysis we used a simple multiplicity trigger where 8 or more DOMs
 recorded a light deposition within a 5 $\mu s$ time window (SMT8).  Most of
 the events which are selected by this trigger are composed of muons produced
 by Cosmic Rays in the atmosphere above the detector (about 2.2 kHz at
 trigger level in the 79-string configuration). These events enter the
 detector only from above since muons produced in the opposite hemisphere of the atmosphere are absorbed by the Earth. Only about one in 10$^{6}$ recorded
 events is induced by an atmospheric neutrino which can reach IceCube from any
 direction. The goal of all further event selections is to increase the ratio
 of the neutrino signal from astrophysical sources with respect to the muon
 background. Key elements of the selection of the neutrino candidates are the
 reconstructions of the event direction and of the deposited energy. Only
 high quality reconstructed events are selected in order to strongly reduce
 the background of downward going muons that are mis-reconstructed as upward
 going. Moreover, since in many signal scenarios the signal is expected to
 have higher energy than the atmospheric background, the estimated energy can be
 used to suppress the low energy muon background. For instance, the first
 order Fermi acceleration mechanisms in SNR shocks predicts a neutrino power
 law spectrum of E$^{-2}$ while the atmospheric neutrinos have a differential
 spectrum in energy above 100 GeV that goes as E$^{-3.7}$~\cite{Atmos}.

\begin{figure}[!t]
  \vspace{5mm}
  \centering
  \includegraphics[width=\textwidth]{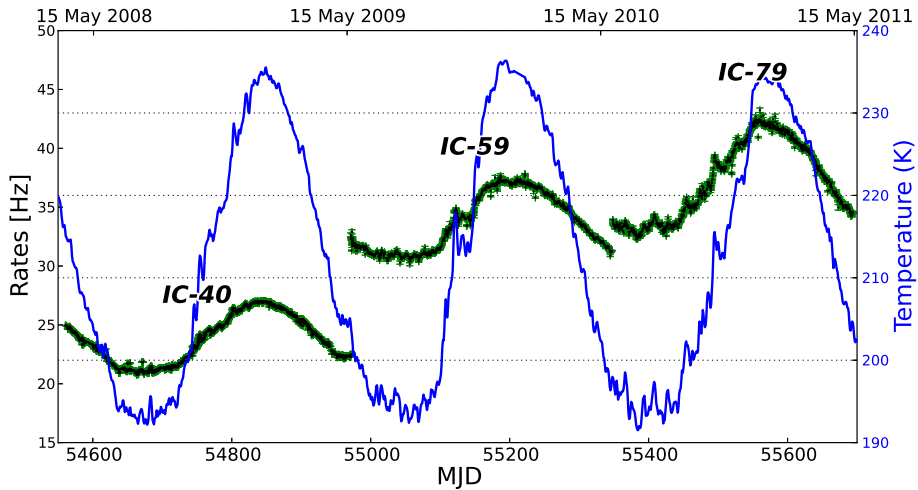}
  \caption{IceCube event rates for the three periods at the muon filter level as a function of the modified Julian date. The correlation with the effective temperature of the South Pole atmosphere is also shown. The effective temperature is a weighted average which takes into account the temperature and pressure at different levels of the atmosphere as well as the attenuation length of the pion as defined in~\cite{temperature}.}
  \label{fig:rates}
 \end{figure}

 A significant part of the background reduction is performed on-line at the
 South Pole (L1 filter) where first cuts on the quality of the reconstruction
 of up-going events are applied and high-energy down-going muons from the
 northern hemisphere are selected. This filtering of events is
 designed to serve a large variety of different muon neutrino searches by
 maintaining a high signal efficiency. This reduction of atmospheric muon
 background is sufficient to send the remaining data off-site by satellite
 where they undergo further processing (L2 filter). Fig.~\ref{fig:rates}
 shows the data rate of each run as function of the modified Julian date for
 one of the data streams of the L2 filter, the muon filter. Also shown is the South Pole
 atmospheric temperature. As can be seen, the atmospheric muon rate is strongly correlated with the atmospheric density which depends on the temperature.

 Most track reconstructions performed at the South Pole are likelihood based
 with the exception of $linefit$, which is an algorithm used as a seed for
 more precise and CPU intense reconstructions to follow. These
 likelihood-based fits use the photon arrival time distribution for track
 reconstruction~\cite{AMANDARec}. The multi photoelectron (MPE) likelihood
 function, which uses time and amplitude information of the PMT pulses, is
 applied after several iterations of the single photoelectron (SPE) likelihood
 fit that uses only the pulse leading edge time.  The energy estimation is
 performed after the track reconstruction since the muon direction
 information is used by the energy reconstruction algorithm. The muon energy
 proxy described in \cite{IC40} was used in all three years of data of this analysis together with a more recently developed algorithm described
 in \cite{te} as alternative energy estimator for the data collected with the
 79-string configuration.

\section{Event selection}
\label{sec3}
The first order background rejection of
the on-line filter is not sufficient for high-level data analyses. Up-going, high-energy neutrino candidates
can be selected from the data by rejecting events with a poor reconstruction
since they are more likely to be down-going muons, by removing mis-reconstructed
events with multiple muon tracks and by suppressing events with low energies. In the southern sky,
it is not possible to reject the muon background based on the reconstruction
quality of the events, and the most energetic events are selected
instead to improve the ratio of signal to background events~\cite{Robert, IC40}. Reflecting the different
detector geometries and the general improvement in the muon track reconstruction
and identification of muon background events, we used a different event selection
for each of the three periods of data used in this work.

The ability to observe a neutrino point source depends on the expected number
of background events, the observable number of neutrinos for a given source
strength, the energies of these events and the angular resolution. The discovery potential,
defined as the flux needed to make a 5$\sigma$ discovery in 50\% of an
ensemble of pseudo-experiments with a simulated signal of this strength,
captures all these aspects and was used as the main figure of merit to
optimize the event selections. Diffuse shock acceleration leads to power-law
spectra with a spectral index around 2~\cite{FermiAcc,FermiAcc2}, and
neutrinos originating in CR interactions near the source are expected to
follow a similar spectrum. We thus used an E$^{-2}$ spectrum as our main
benchmark model. Several galactic $\gamma$-ray sources have energy spectra
with energy cut-offs at a few TeV~\cite{Julia}, supporting the idea that
galactic neutrino spectra may present cut-off spectra as well. We therefore
also took softer neutrino spectra into account. These softer spectra were
modeled by larger spectral indexes (e.g. 2.7 or 3) and/or by exponential
energy cut-offs.

\subsection{IceCube 40-strings data sample}
During the IC-40 period, IceCube was
active and taking data more than 99\% of the time, and 92\% of the data
were used after selecting periods of stable detector operation. The data used after this selection
correspond to a live-time of 375.5 days. 
The event selection for the point source analysis of the IC-40
data was obtained by cuts on a number of
well-understood and powerful variables and is described in detail in
Ref.~\cite{IC40}. In the southern sky, events were selected with a cut on the reconstructed energy 
of the event which was parameterized as a function of the reconstructed declination. 
The final sample of events obtained from the IC-40 configuration contained a total
of 36,900 events: 14,121 from the northern sky and 22,779 from the southern
sky.

\subsection{IceCube 59-strings data sample}

The data from the IC-59 configuration correspond to a live-time of 348.1 days.
The rate of the SMT8 trigger was of the order of 1.5 kHz, and the on-line muon filter rate
was a factor two higher than in the previous configuration as can be seen in Fig.~\ref{fig:rates}.

\begin{figure}[t]  

\begin{tabular}{c c}
\includegraphics[width=0.55\textwidth]{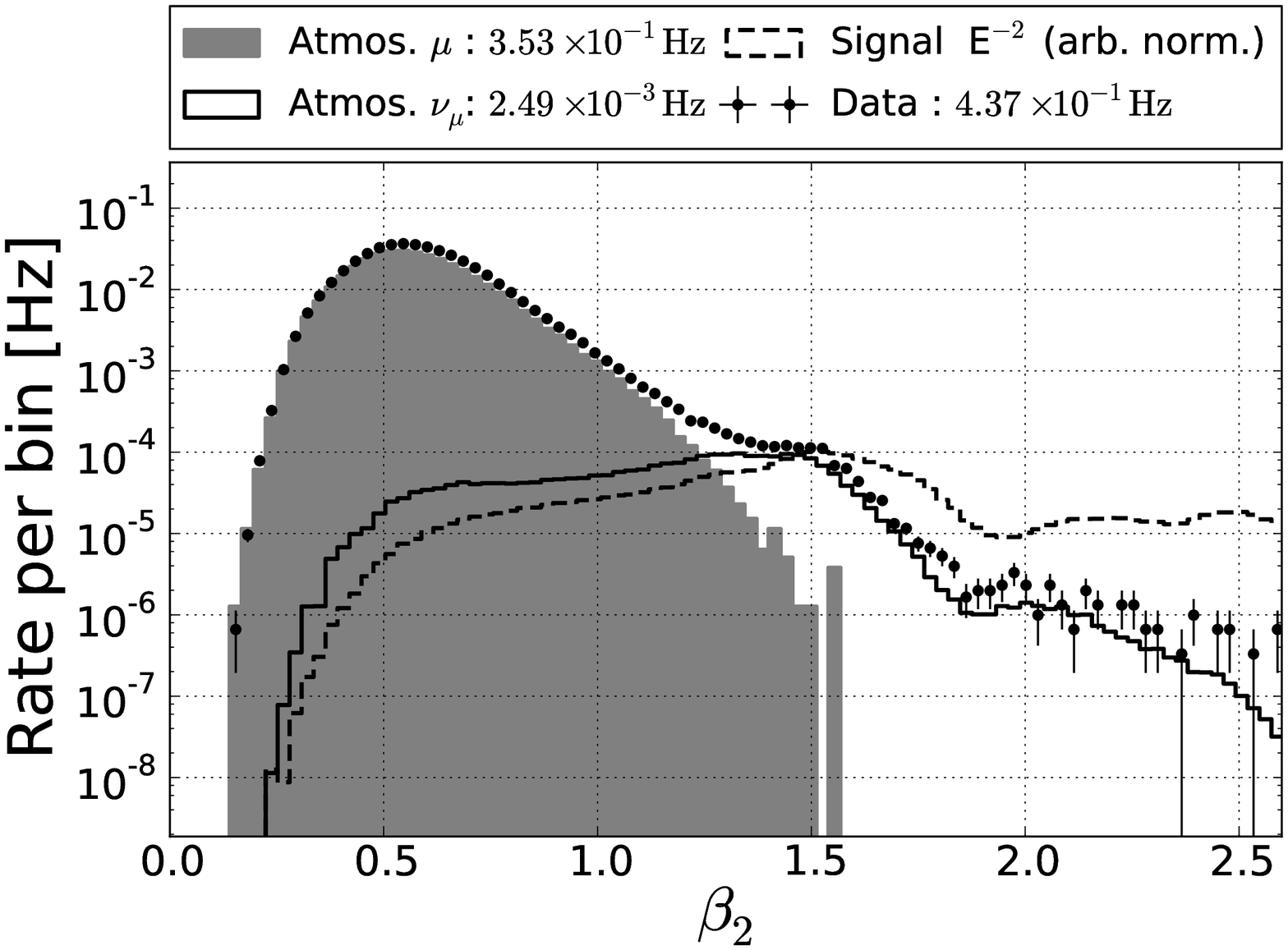}
\hspace{-1cm}
\includegraphics[width=0.55\textwidth]{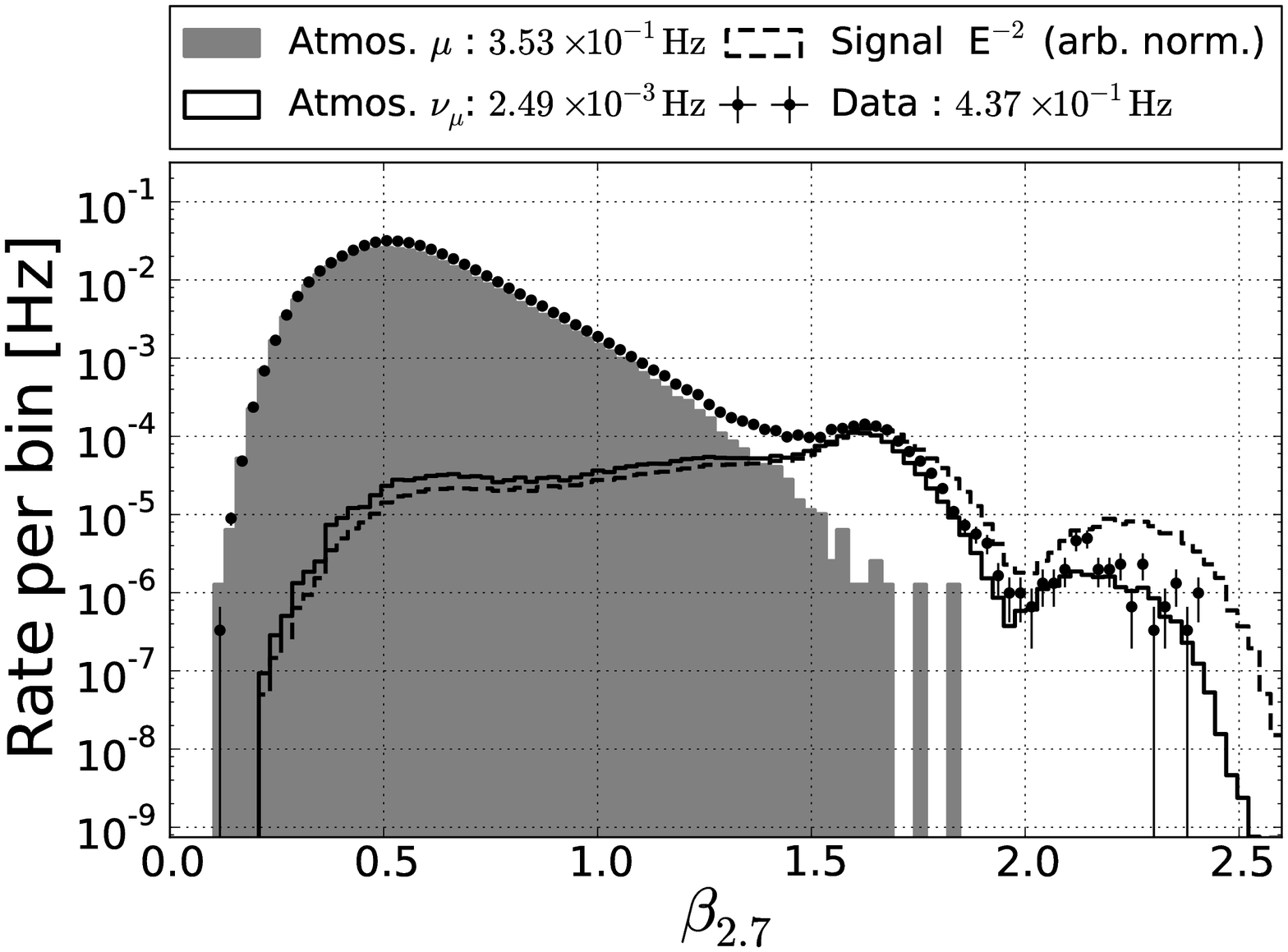} 
\end{tabular}
\caption{\label{fig:IC59BDT} Distribution of the BDT score for the ensemble of trees trained with an E$^{-2}$ spectrum (left) and an E$^{-2.7}$ spectrum (right) as signal. The data shown here are from the IC-59 configuration and examples for a signal distribution are shown with an arbitrary normalization.}
\end{figure}

It was shown in Refs.~\cite{ICAMA} and \cite{Crab} that a higher efficiency for up-going
neutrino events with energies below 10 TeV can be achieved with multi-variate
approaches without compromising the discovery potential for neutrino sources with hard
energy spectra. In the IC-59 data sample, we used Boosted Decision Trees
(BDTs)~\cite{BDT} to this end. BDTs are widely used in two-class
classification problems where a larger set of weakly discriminating variables
is available~\cite{Stats} and are thus well-suited for the selection of
neutrino events from the IceCube data.

The multivariate cuts were based on twelve observables with a high discriminating power
between signal and background. 
We used 10\% of the atmospheric muon dominated data as a background model for the BDT training. Any possible
astrophysical signal contributes only a very small fraction to the data at
this level. The observables were selected such that their correlation in the background-dominated data sample
were below 50\%. The signal was modeled with Monte Carlo simulation. Two
different signal energy spectra were considered: an E$^{-2}$ spectrum, and
one with a neutrino spectrum of E$^{-2.7}$ to account for softer neutrino
spectra. Additionally, the reconstructed track was required to be within
0.5$^{\circ}$ of the simulated direction in order to train the BDT with only
well-reconstructed events.

For computational reasons, the observables were split in two sets of eight and four
variables, and a BDT was defined for each set separately. The final selection
was based on a combination of the two BDT scores. Fig.~\ref{fig:IC59BDT}
shows the distribution of the combined BDT scores
obtained by the training with the two different signal spectra for data and a simulated
neutrino signal as well as for the simulated atmospheric muon and neutrino
backgrounds.

Events in the southern Hemisphere were selected with a cut on the reconstructed energy. The strength of the cut was varied
as a function of the declination. In addition, we used the veto capability of the surface array IceTop \cite{IceTopVeto} to
reduce the muon background. Atmospheric muons are accompanied by extended air showers
which can produce early hits in the IceTop surface array. The veto is defined
by counting the number of detected photo electrons in IceTop within a time
window around the expected arrival time of the shower front in the surface
detector. 
In the IC-59 event selection, the IceTop veto
was used for events with reconstructed declinations between $-90^{\circ}$ and $-40^{\circ}$.
The best veto efficiency is
expected for events with high energies, heavy primaries, vertical directions
and a shower axis close to the IceTop detector. Fig.
\ref{fig:icetopveto_79} shows the veto capability of the IceTop surface array
using atmospheric muon dominated data from IC-79. The IceTop veto allows us to reject background with
~99\% efficiency in the vertically down-going region without losing signal
neutrino efficiency ( $\lesssim 1$\%).

The final data sample for the IC-59 configuration has a total number of
107,569 events, among which almost 2/3 come from the southern sky. The rest
are neutrino candidates in the northern sky.

\subsection{IceCube 79-strings data sample}
As illustrated in Fig.~\ref{fig:layout}, the IC-79 configuration had almost the final
volume of the full IceCube detector. With the largest detector size among the configurations discussed
here, the background from coincidences of two or more atmospheric muons
within the same read-out window is more abundant than
in the previous ones. At the same time, the number of neutrino events in
coincidence with an atmospheric down-going muon increased as well. 
We applied a topological hit clustering based
on the spatial and temporal separation of recorded PMT signals to separate
neutrinos from coincident muons. In addition to the reconstruction of the
full event, we applied the same reconstruction to up to three topologically
connected subsets of hits. Among these and the original reconstructed track,
only the most likely neutrino candidate was selected. If only
the reconstruction of the full hit information passed the cuts but none of
the subsets, the events were rejected to improve the background suppression.
A visual inspection of more than 50 events at
the final selection level showed that the topological splitting of events
allowed us to select additional, high-quality neutrino events from which a
coincident muon contamination was removed.

\begin{figure}[!t]
  \vspace{5mm}
  \centering
  \includegraphics[width=4.in]{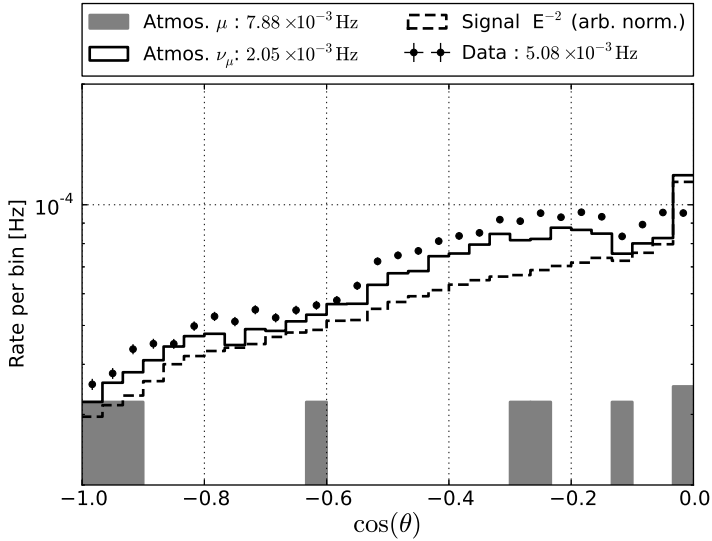}
  \caption{Zenith distribution at final cut level for the up-going neutrino selection. Points represent data. The dashed line is a benchmark E$^{-2}$ astrophysical neutrino signal normalized to the all-sky atmospheric neutrino rate shown to illustrate how an hypothetical signal distributes in zenith. The solid line is the simulated atmospheric neutrino contribution and the filled histogram shows the simulated contribution of mis-reconstructed atmospheric muons after all cuts estimated.}
  \label{fig:zenith79A}
 \end{figure}

Two different high-level event selections have been developed, which we denote sample A and B in the following. While all the results presented in this article have been obtained on sample A, we used sample B for cross-checks and validation of the point source analysis.

\begin{figure}[t]
  \centering
  \includegraphics[width=2.5in]{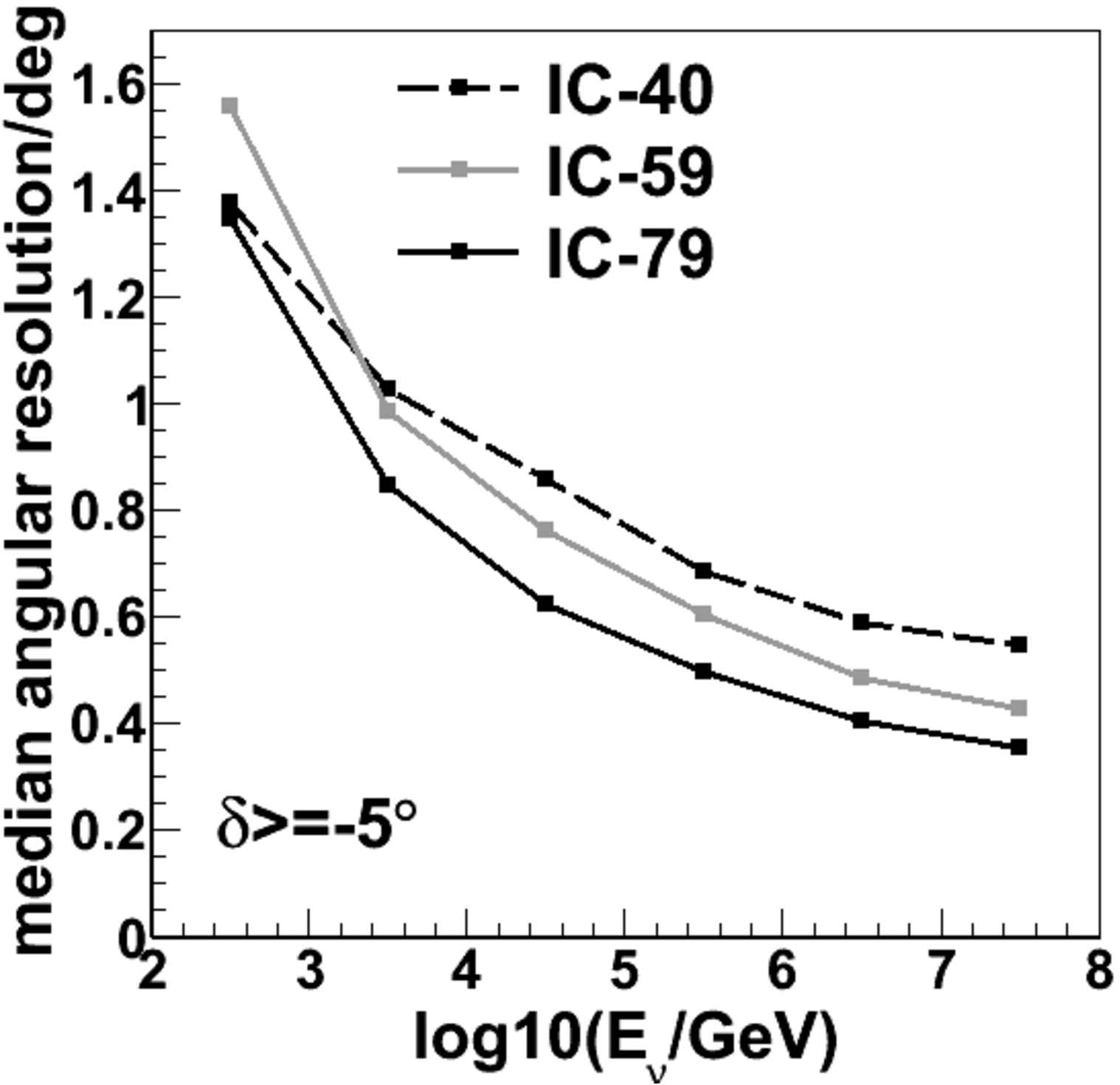} 
  \includegraphics[width=2.5in]{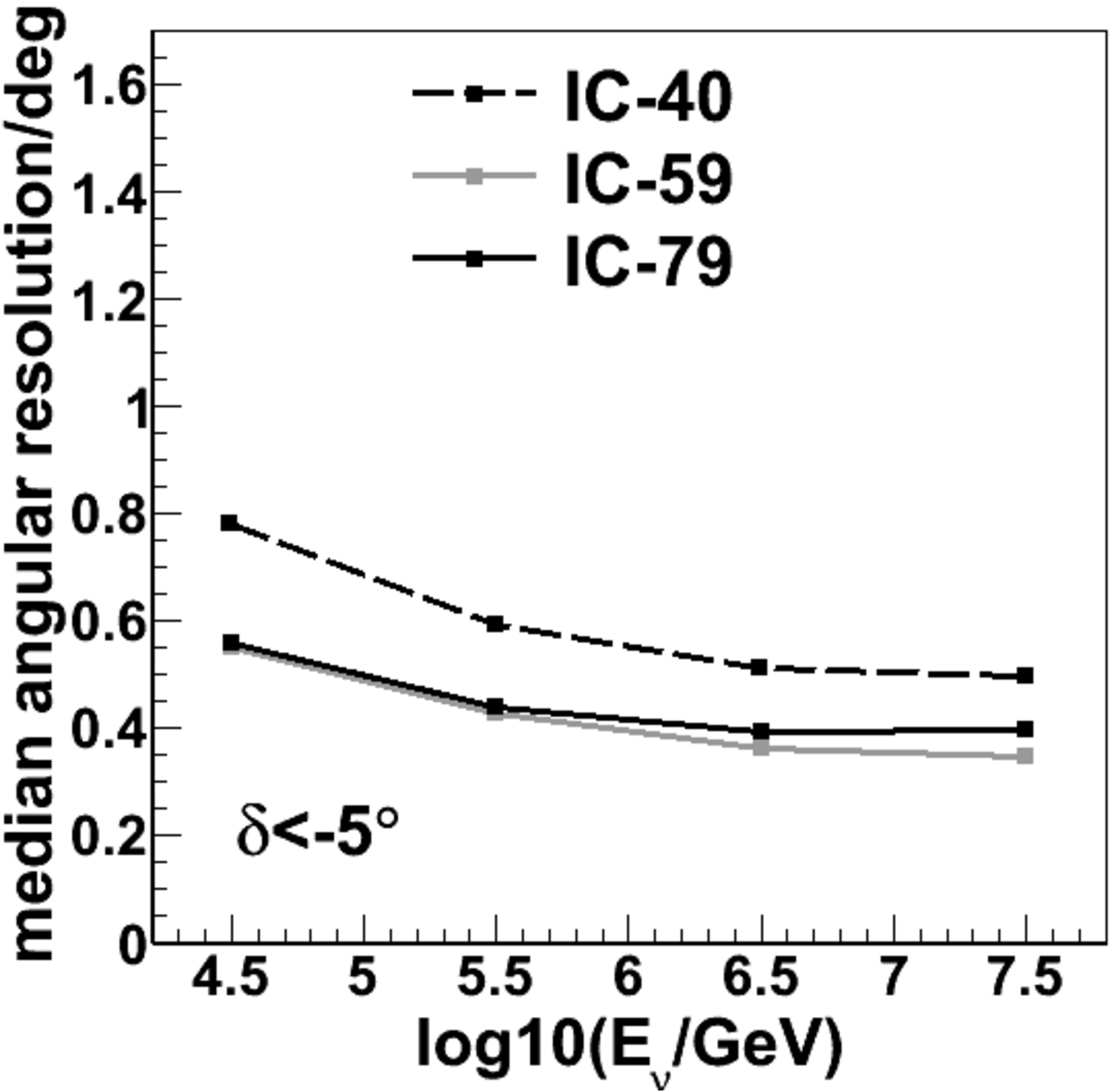} 
  \caption{Neutrino angular resolution defined as the median of the point spread function of the true neutrino direction and the reconstructed muon direction for northern (right) and the southern sky (left) at analysis level.}
  \label{fig:angular_resolution}
 \end{figure}

For both event samples, we used a combination BDTs for the event selection in the northern sky
and divided this area into two regions, a horizontal and a vertical one. For sample A (B), we 
considered events within $-5^{\circ} \le \delta \le +40^{\circ}$ ($0^{\circ} \le \delta \le +35^{\circ}$) and events within
$+40^{\circ} \le \delta \le +90^{\circ}$ ($+35^{\circ} \le \delta \le +90^{\circ}$) separately.
The two bands are characterized by different expected
signal energy spectra due to the absorption of high-energy neutrinos in the
Earth and by different distributions of the background.

For sample A, seventeen observables were selected for the BDT based on their discrimination
power between a neutrino signal and the muon background. They were split into two sets of nine and eight variables each. In addition, we asked for the background and signal correlation
coefficients between any two variables in the same set to be below 50\%. A
number of variables with less discrimination power were included since BDTs are robust against
the inclusion of weak variables. The final cut parameter was defined
by a combination of the two BDT scores for each zenith region.

In sample B, only highly discriminating variables were included in the BDTs and the event
selection used a different number of variables in the horizontal
region where the signal is dominated by higher energy events. Nine variables were selected for the vertical region and fifteen
variables for the horizontal region. No requirement
was applied to limit the correlations between the variables, allowing us to
use all selected variables in a single BDT in each region. Eight of the
observables used for the BDT sample B were also used in sample A.
 
As in the IC-59 selection we trained the BDTs with two different signal
spectra, using again the E$^{-2}$ spectrum as a benchmark for hard
spectra. As representative of a soft spectrum, we used an E$^{-2.4}$ spectrum
with a cutoff at 7 TeV for sample A and an E$^{-2.7}$ spectrum for sample
B.

The final selections were optimized to provide the best discovery potential for E$^{-2}$ neutrino fluxes. At the same time, we aimed to achieve a near-optimal discovery potential for softer spectra by adding additional lower energy events. We also paid special care in having a smooth transition in the event rate between the two declination regions. Fig.~\ref{fig:zenith79A} shows the zenith distribution for the up-going event selection in sample A. As can be seen, it is fully dominated by atmospheric neutrinos and only a small fraction of mis-reconstructed down-going atmospheric muons survive after the event selection.

\begin{figure}[!t]
  \vspace{5mm}
  \centering
  \includegraphics[width=4in]{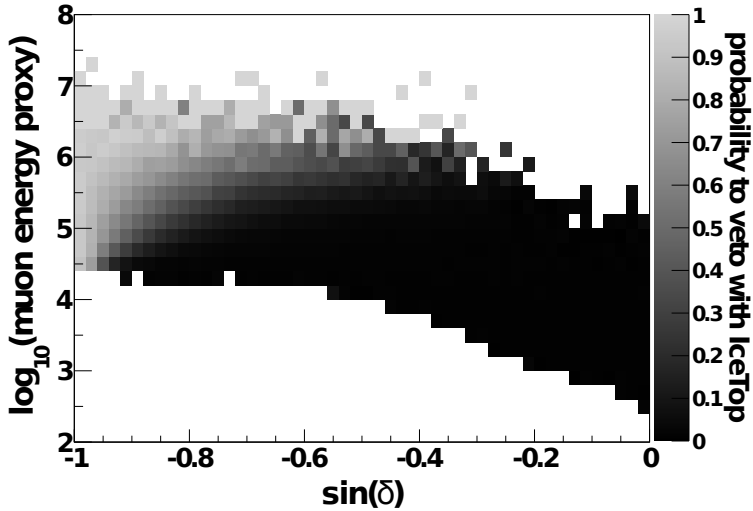} 
  \caption{IceTop veto capability  as a function of the sine of the declination, $\delta$, of the detected muon and the estimated energy of the detected muon in the detector based on the data from IC-79. The events used for this visualization passed an intermediate set of event selection criteria but are still dominated by atmospheric muons. Events with 3 or more veto hits within a time window of $\pm 1000$ ns around the expected shower front arrival time are rejected. 10\% of the experimental data were used for this plot and the white areas correspond to regions where no event was observed.}
  \label{fig:icetopveto_79}
 \end{figure}

\begin{figure}[!t]
  \vspace{5mm}
  \centering
  \includegraphics[width=4.in]{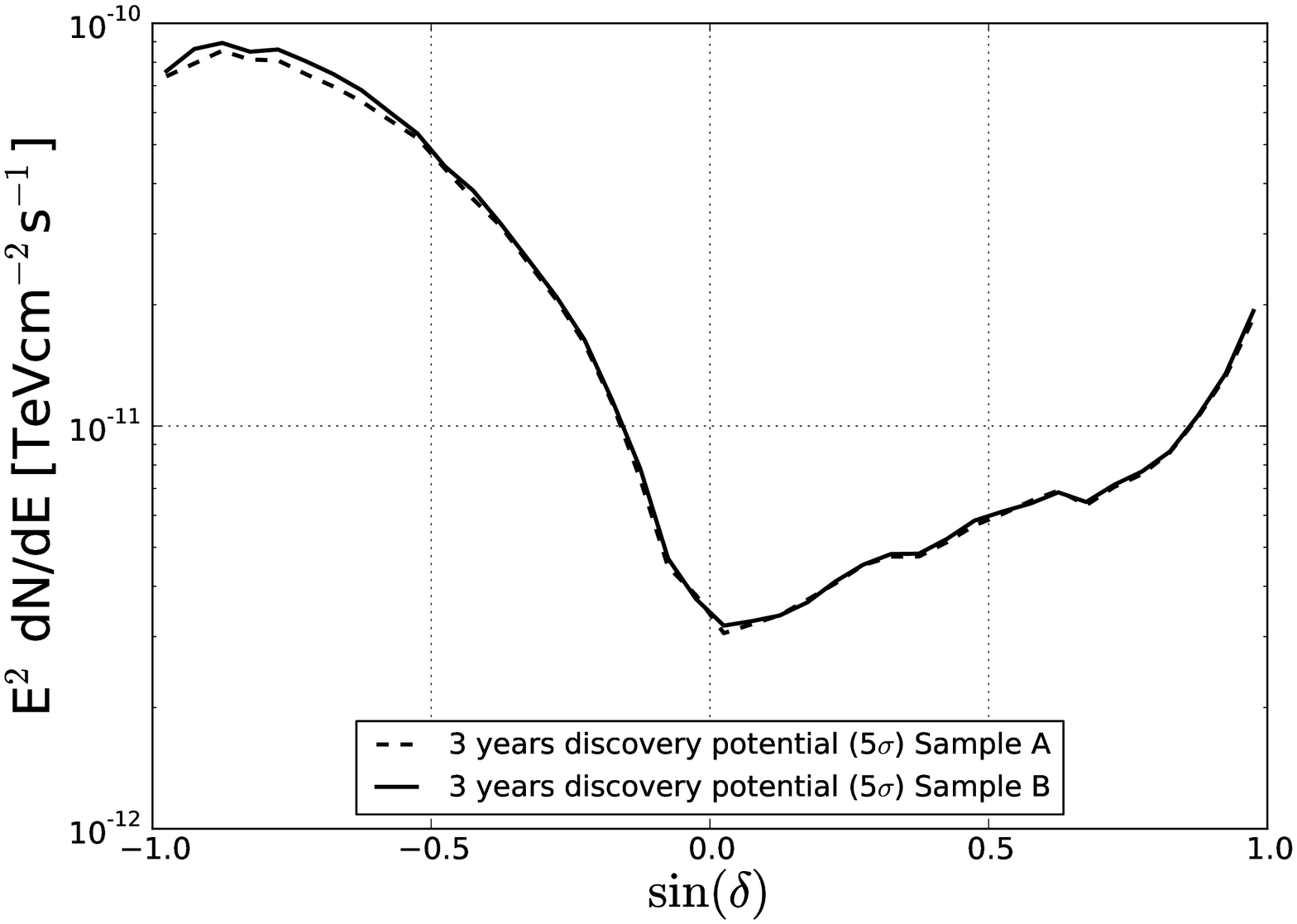}
  \caption{Discovery potential for point sources at 5$\sigma$ confidence
    level for an E$^{-2}$ spectrum as a function of declination for the 3
    years of IceCube when sample A (dashed) or sample B (solid) is used for the
    IC-79 data.}
  \label{fig:comparison}
 \end{figure}

For sample A, we extended the cuts from the up-going region to the southern
hemisphere by using the same set of cuts and a retrained BDT as an intermediate
event selection. In addition, an angular uncertainty estimator was required to be smaller than
2$^{\circ}$. We also applied a veto based rejecting events with three or more veto hits in the surface array IceTop. The probability to veto events by accidental coincidences was estimated by using experimental data from an off-time window where no correlated signal
in IceTop is expected. It is below 1\% at every
declination and energy. The probability to veto a background event is shown in Fig.~\ref{fig:icetopveto_79}. The
background rejection power is above 90\% for high-energy, vertical
down-going muons. To further decrease the rate of accidental
coincidences, we applied the IceTop veto cut only in those regions of the
energy-declination space where it is most efficient. In this way, the
accidental veto probability is much smaller than 1\% and its effect on the
signal efficiency can be neglected. The effect of the IceTop veto in the IC-79 and IC-59 event selection is visible in
Fig.~\ref{fig:comparison}. For very vertical down-going events ($\sin \delta < -0.85$) where the veto is most efficient there is a decrease in the discovery flux. Finally, a declination-dependent
energy cut was used to select a constant number of events per solid angle
and to provide a smooth transition from the northern to the southern
hemisphere.

For sample B, a simple energy cut depending on the declination was applied
to select a constant number of events per solid angle. The same, soft IceTop veto as above was used to reject part of
the down-going atmospheric muon background at the very vertical zenith
angles. A study performed on this sample indicated that no significant gain
in the discovery potential could be achieved by selecting a larger number of
events in the Southern Hemisphere. 

Sample A contains 109,866 events of which 50,857 are coming from the northern sky and 59,009 are located in the southern sky. 

The two samples yield the same discovery potential for steady, point-like
neutrino sources both for hard (represented by an E$^{-2}$ neutrino signal
spectrum) and soft (represented by an E$^{-3}$ neutrino signal spectrum)
neutrino spectra at every declination (see Fig.~\ref{fig:comparison}). The
differences are smaller than the statistical uncertainty of the
estimation. This is a confirmation of the
validity of the independent BDT selections.

Sample B has a slightly larger effective area for events at lower energies
than sample A at the cost of a higher muon contamination in the northern
hemisphere. Considering events with reconstructed declinations above
0$^{\circ}$, we observe that the events which are contained in both samples
make up 75\% of the events in sample A and 67\% of the events
in sample B. The difference in the percentages reflects the smaller number
of events in sample A. The overlap between the two samples is larger for
events with small angular uncertainties, rising to 81\% of the events in
sample A and 90\% of sample B being contained in both samples if events with
angular uncertainty estimates smaller than 0.5$^{\circ}$ are
considered. Thus, the probability for a more
signal-like, well-reconstructed event to be in both samples is higher than
the corresponding probability for an event with a poor reconstruction. A
visual inspection of the hit patterns of a subset of the events confirms
that the contribution of background events from atmospheric muons is smaller
in the group of events which are in both samples. In particular, we have
visually checked the hit patterns of more than 100 up-going events which are
in both samples and have an angular uncertainty estimate smaller than
0.5$^{\circ}$; all of these were well-reconstructed up-going,
i.e.\ neutrino-induced events.

The overlap of the two event samples in the southern sky is much smaller
than in the northern sky. In the region from declination $-90^{\circ}$ to
$0^{\circ}$, we observed that 38\% of the events in sample A are also
contained in sample B and that 27\% of the events in sample B are also in
sample A. The fraction of events common in both samples increases for smaller
angular uncertainty estimates. The smaller overlap is expected. 
The event selection in A disfavors events with very large energy losses with respect to the event selection
B. Moreover, both event selections apply a filtering of different strength
before the energy cut is applied on the steeply falling spectrum. Moving the
strength of the energy cut at any declination by a small amount will
decrease the overlap between two event selections significantly.

\begin{figure}[h!]
\hspace{-1cm}
\begin{tabular}{c c}
\includegraphics[width=0.55\textwidth]{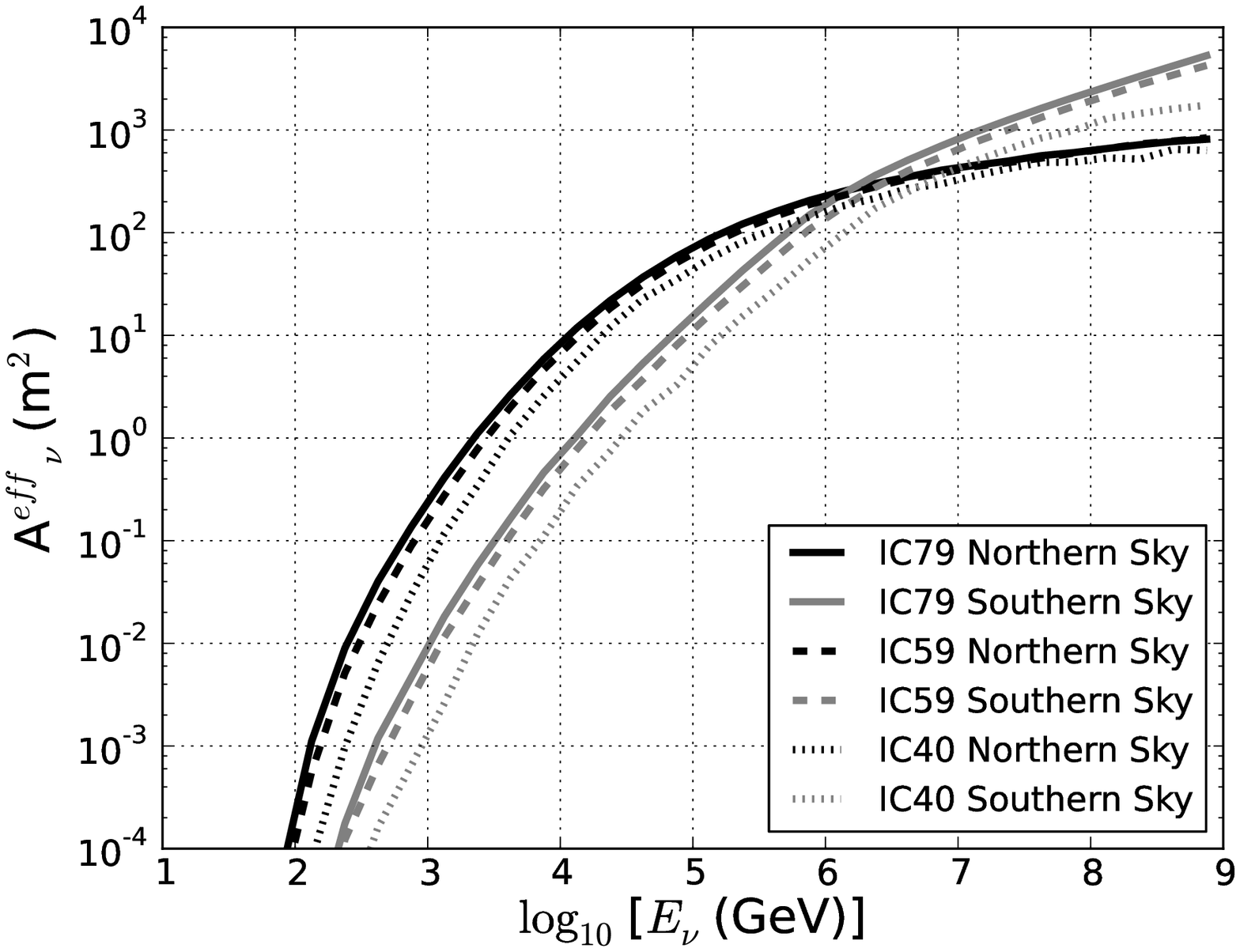} 
\includegraphics[width=0.55\textwidth]{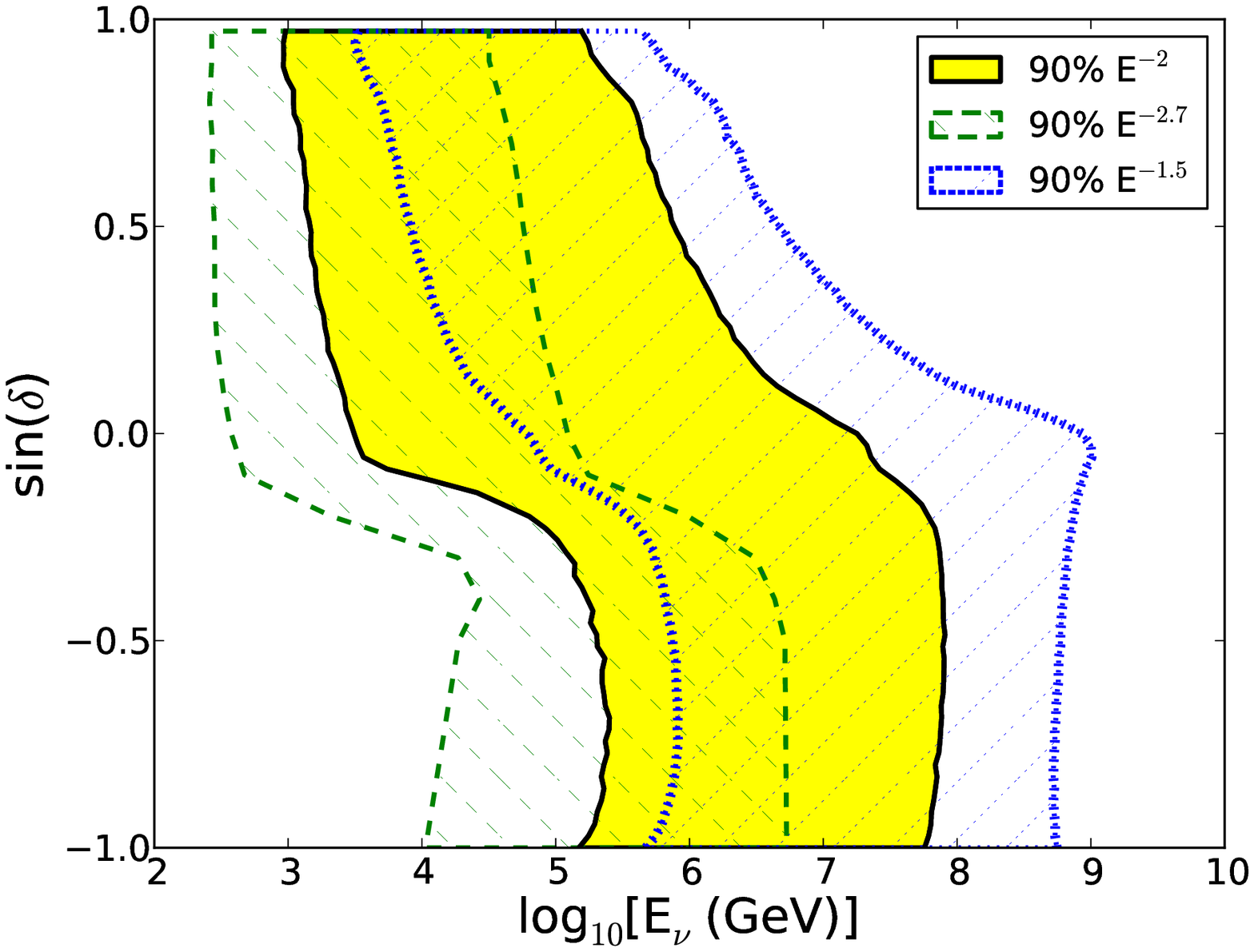} 
\end{tabular}
  \caption{Left: Solid-angle-averaged muon neutrino effective area for the
    three event selections corresponding to three IceCube configurations for
    both the northern and southern skies. Right: 90\% central signal containment region for
    three different power-law neutrino spectra as a function of
    declination for the three configurations combined.}
  \label{fig:aeff_ic40_ic59_ic79}
 \end{figure}

\begin{figure}[h]
\centering  
\includegraphics[width=0.75\textwidth]{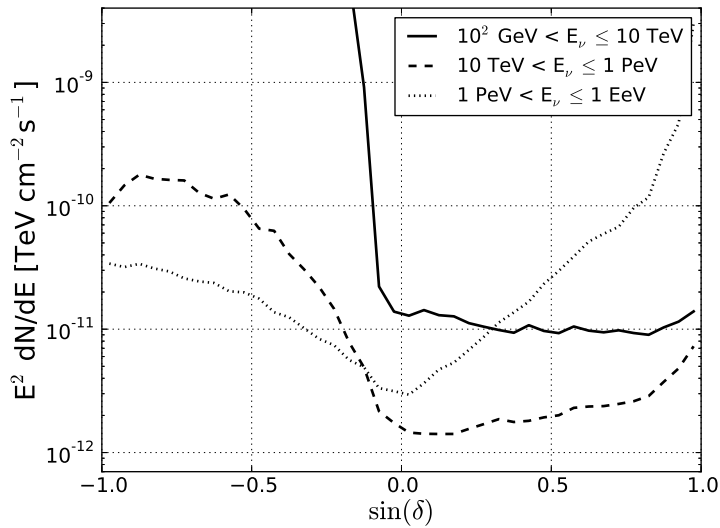} 
\caption{Sensitivity for muon neutrino flux for a E$^{-2}$ spectrum for a
  90\% C.L. as a function of declination combining the three years of data averaged over right ascension. The three different lines indicate
  three different energy ranges.}
  \label{fig:sensi_energies}
 \end{figure}

 Table~\ref{tab:lifetimes} summarizes the live-time, the estimated rate of
 atmospheric neutrinos and the number of up-going and down-going track events
 in the three different configurations using sample A for the IC-79 configuration. 
 Fig.~\ref{fig:angular_resolution} shows the neutrino angular resolution in each of the three data samples.
 In the northern sky, the best angular resolution is observed in IC-79. The lowest energy bin in the IC-40 sample has a better resolution than in the IC-59 sample because of the stricter event filtering applied in this energy range. The southern sky selection of the IC-59 sample applies stronger cuts than the IC-79 selection, leading to a slightly better angular resolution in this range. The effective area for muon neutrinos, $A_{eff}^{\nu}$, after the final event selection for the three configurations of IceCube is shown in Fig.~\ref{fig:aeff_ic40_ic59_ic79} left. For the tabulated data of this figure see the appendix~\ref{AppAeff}. The right panel in Fig.~\ref{fig:aeff_ic40_ic59_ic79} shows the 90\% central signal containment for three power-law neutrino spectrum of E$^{-2}$, E$^{-2.4}$ and
E$^{-1.5}$ using the combination of the three different detector geometries. These regions indicate the sensitivity range of IceCube for different spectral indexes. The effect in sensitivity for point sources of neutrinos can be
clearly seen in Fig.~\ref{fig:sensi_energies}. The dashed line represents
the expected sensitivity at the 90\% confidence level (C.L.) as a function of declination for an
E$^{-2}$ signal in the energy range between 10 TeV and 1 PeV where most of the signal deposition is expected for this spectrum. The dotted and solid
lines show the sensitivity for an E$^{-2}$ in a higher and lower energy
range. As can be seen, IceCube's sensitivity for low energy neutrinos (E $\leq$ 10~TeV) is mostly restricted to the northern sky while at higher energies the sensitivity of the detector is more symmetric.

\begin{table*}[t!]
\begin{center}
  \caption{Summary for three different IceCube configurations for point
    source analyses: The expected atmospheric neutrino rate from MC
    simulation weighted for the model in Ref.~ \cite{Honda} and numbers of
    up- and down-going events at final selection level. The numbers for the
    IC-79 are from sample A. }
\vspace{0.2cm}
\small{
\begin{tabular}{ccccc}
\toprule
no. of strings & live-time [days] & atm. $\nu$s & \# up-going &  \# down-going\\
\midrule 
40  &  376 &  40/day & 14,121 & 22,779 \\
59  &  348 &  120/day& 43,339 & 64,230 \\
79  &  316 &   180/day& 50,857 & 59,009 \\
\bottomrule
\end{tabular}}
\label{tab:lifetimes}
\end{center}
\end{table*}

\section{The likelihood search method}
\label{sec4}

To search for neutrino point sources in IceCube we use an unbinned maximum
likelihood ratio test. This method follows the one described in~\cite{method}
and is extended to combine different detector geometries. It calculates the significance of an excess of neutrinos over the atmospheric 
background by using both the directional information of the events and the energy to separate hard-spectrum signals from the softer
spectra of atmospheric neutrinos and muons. The method models the expected neutrino signal from
a point source in the sky using simulation and since this search
is background dominated its estimate is done using real data.

The signal and background probability density functions (p.d.f.) are a function
of the reconstructed declination and the reconstructed muon energy.

For a data sample of $N$ total events the p.d.f of the $i^{th}$ event in the
$j^{th}$ sample (in our case the IC-40, IC-59 or IC-79 datasets) with reconstructed energy $E_{i}$ and located at an angular distance to the source of $|\vec{x}_i-\vec{x}_{s}|$ is given by:

\begin{equation}
\label{eq:pdf}
P^j_i (|\vec{x}_i-\vec{x}_{s}|, E_i, \gamma, n^{j}_{s}) = \frac{n^{j}_{s}}{N^{j}}\mathcal{S}^{j}_{i} + \left(1- \frac{n^{j}_{s}}{N^{j}}\right)\mathcal{B}^{j}_{i},
\end{equation}

\noindent where $\mathcal{S}^j_i$ and $\mathcal{B}^{j}_i$ are the signal and
background p.d.f. respectively and $n^{j}_{s}$ is the fraction of total
number of signal events, $n_{s}$, that is expected from the corresponding $j^{th}$ sample.

For time integrated searches the signal p.d.f. $\mathcal{S}^{j}_i$ is given by:
 
\begin{equation}
\label{eq:signalpdf}
\mathcal{S}^j_i = S^{j}_{i} (|\vec{x}_i-\vec{x}_{s}|,\sigma_{i}) \mathcal{E}^{j}_{i} (E_i, \delta_i, \gamma),
\end{equation}

\noindent here $S^{j}_{i}$ is the space contribution and depends on the angular uncertainty of the event, $\sigma_i$, and the angular difference between the reconstructed direction of the event and the source. We model this probability as a 2-dimensional Gaussian,

\begin{equation}
\label{eq:space}
S^j_i = \frac{1}{2\pi\sigma^2_i}e^{-\frac{|\vec{x}_i-\vec{x}_{s}|^2}{2\sigma^2_{i}}}.
\end{equation}

\begin{figure}[htpb]
  \vspace{5mm}
  \centering
\begin{tabular}{c c}
  \includegraphics[width=.5\textwidth]{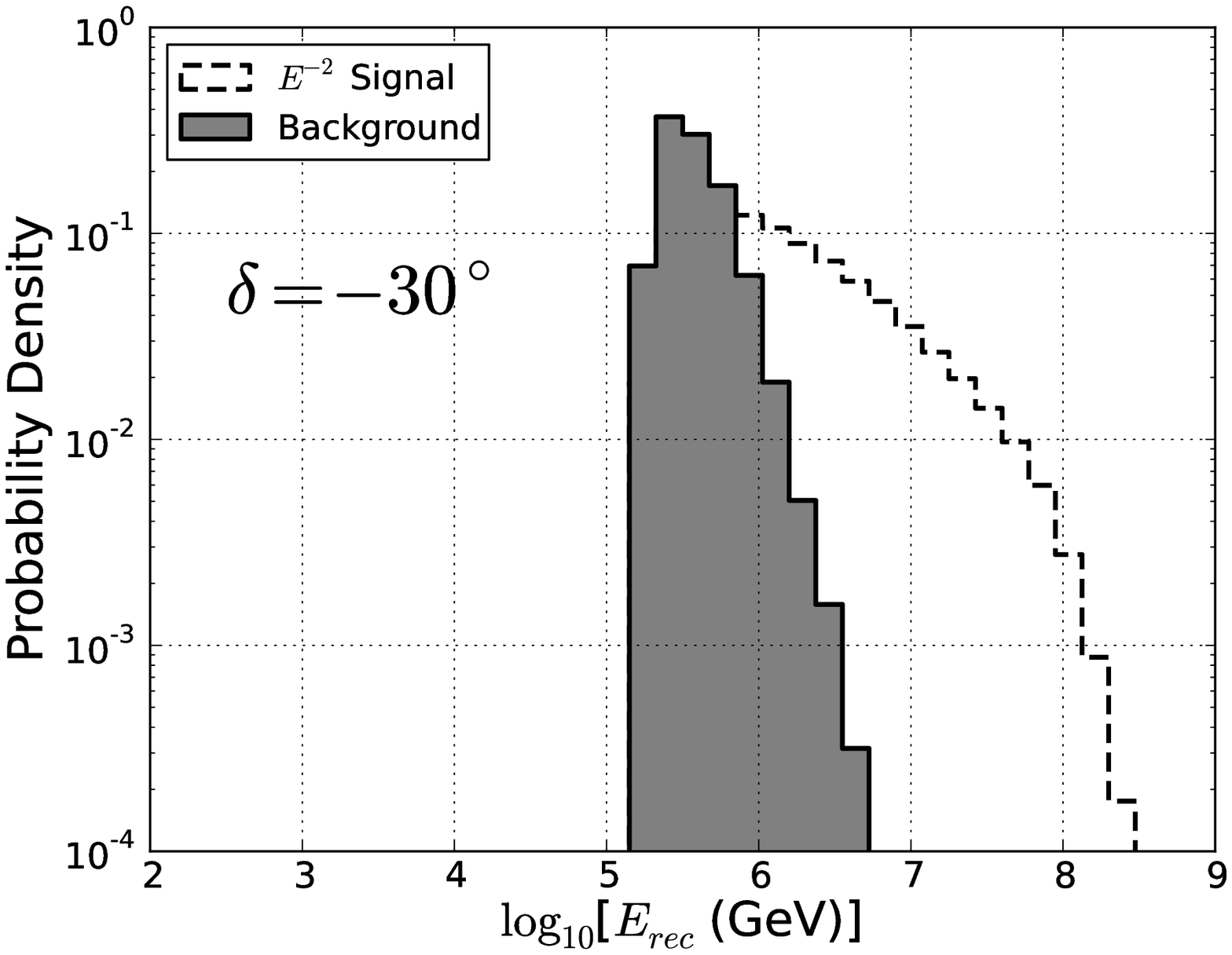} 
  \includegraphics[width=.5\textwidth]{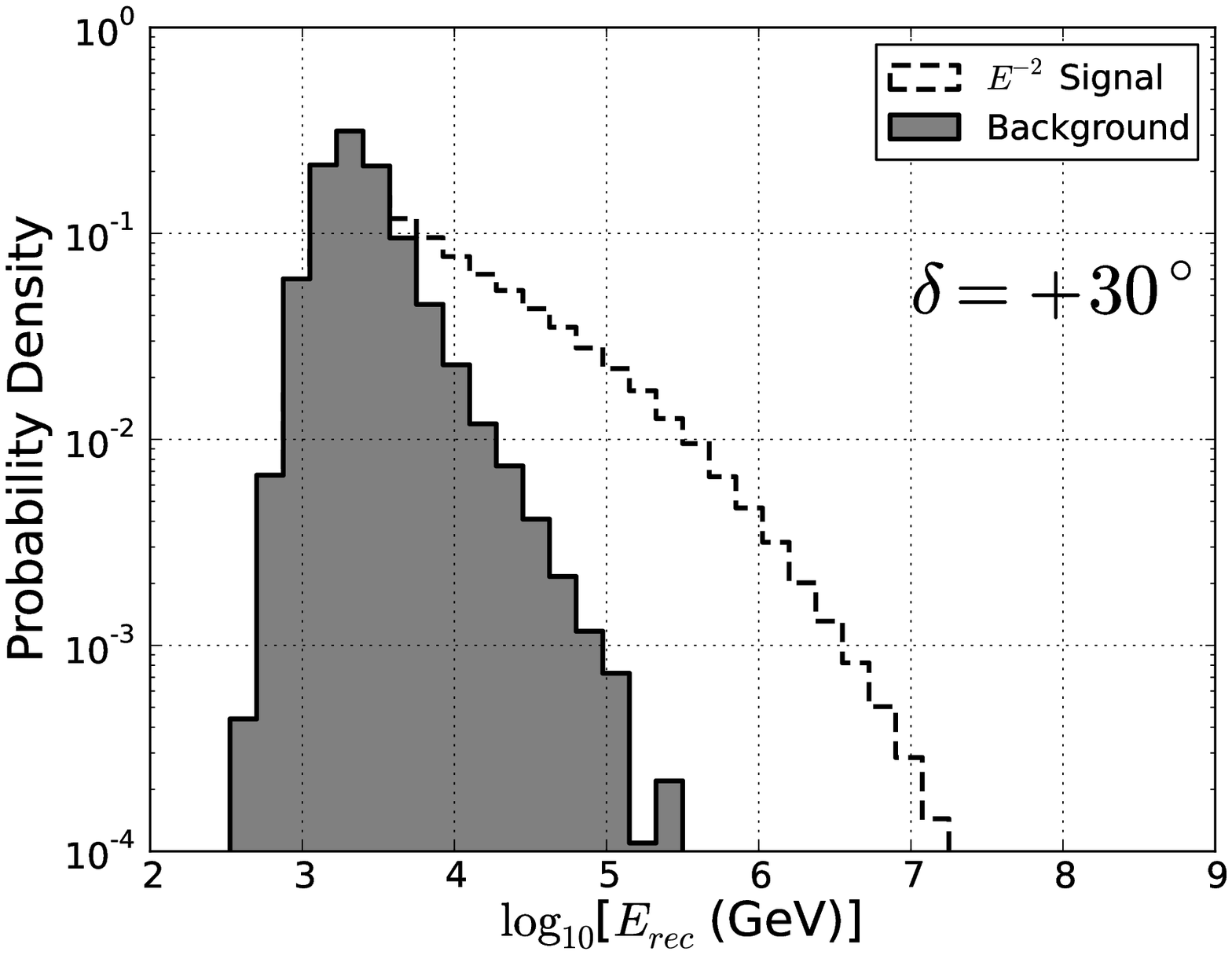}\\ 
 \includegraphics[width=.5\textwidth]{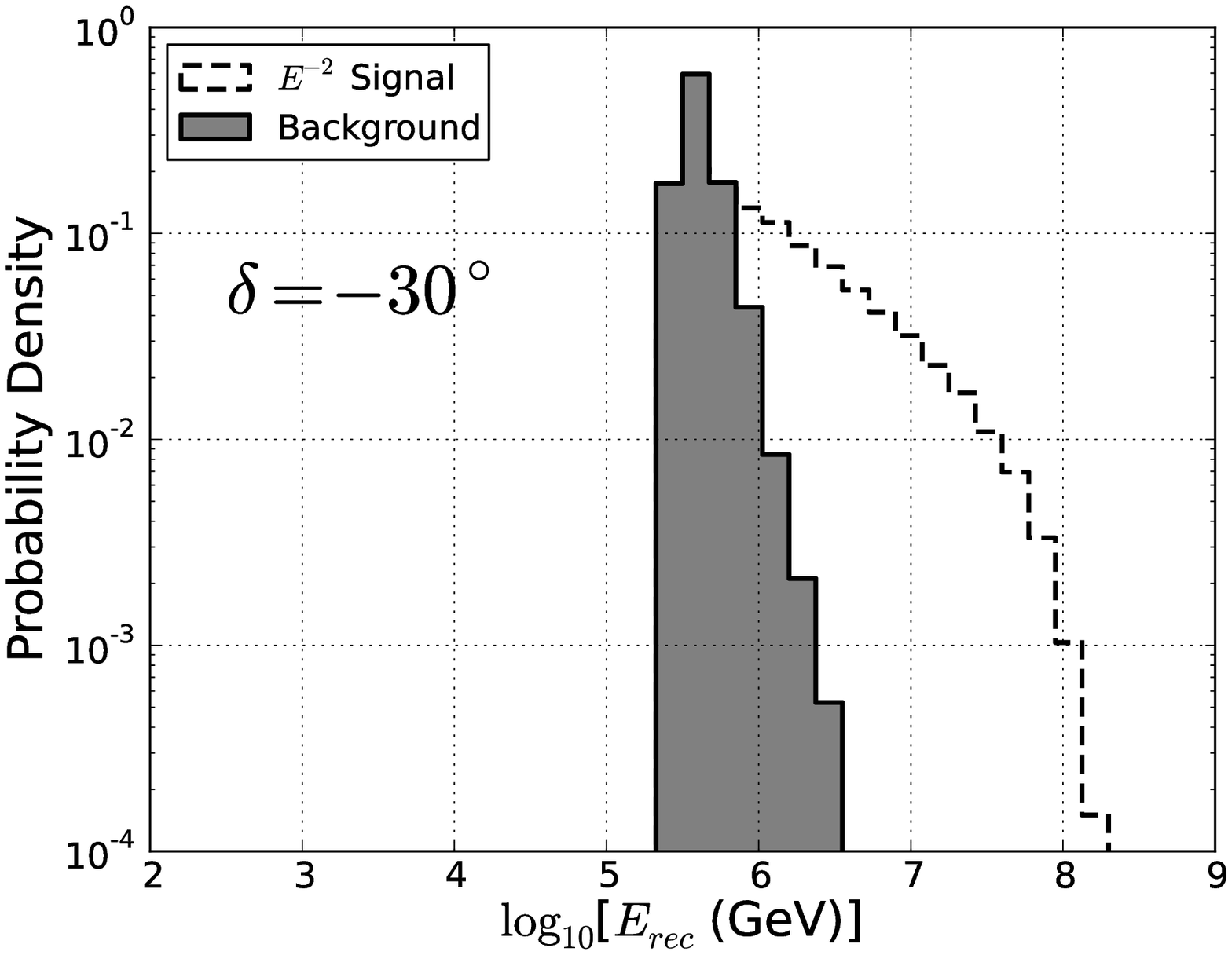} 
  \includegraphics[width=.5\textwidth]{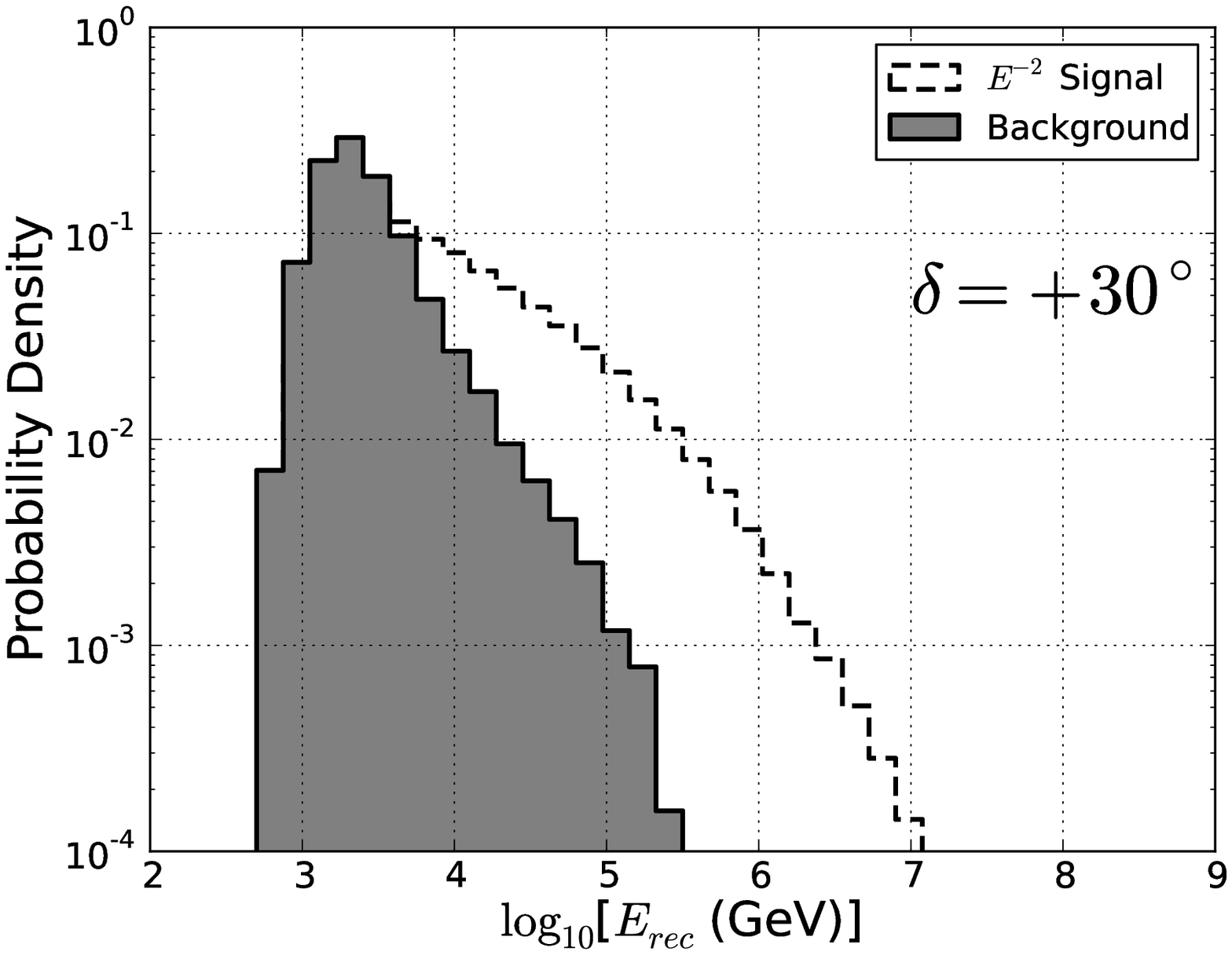}\\ 
 \includegraphics[width=.5\textwidth]{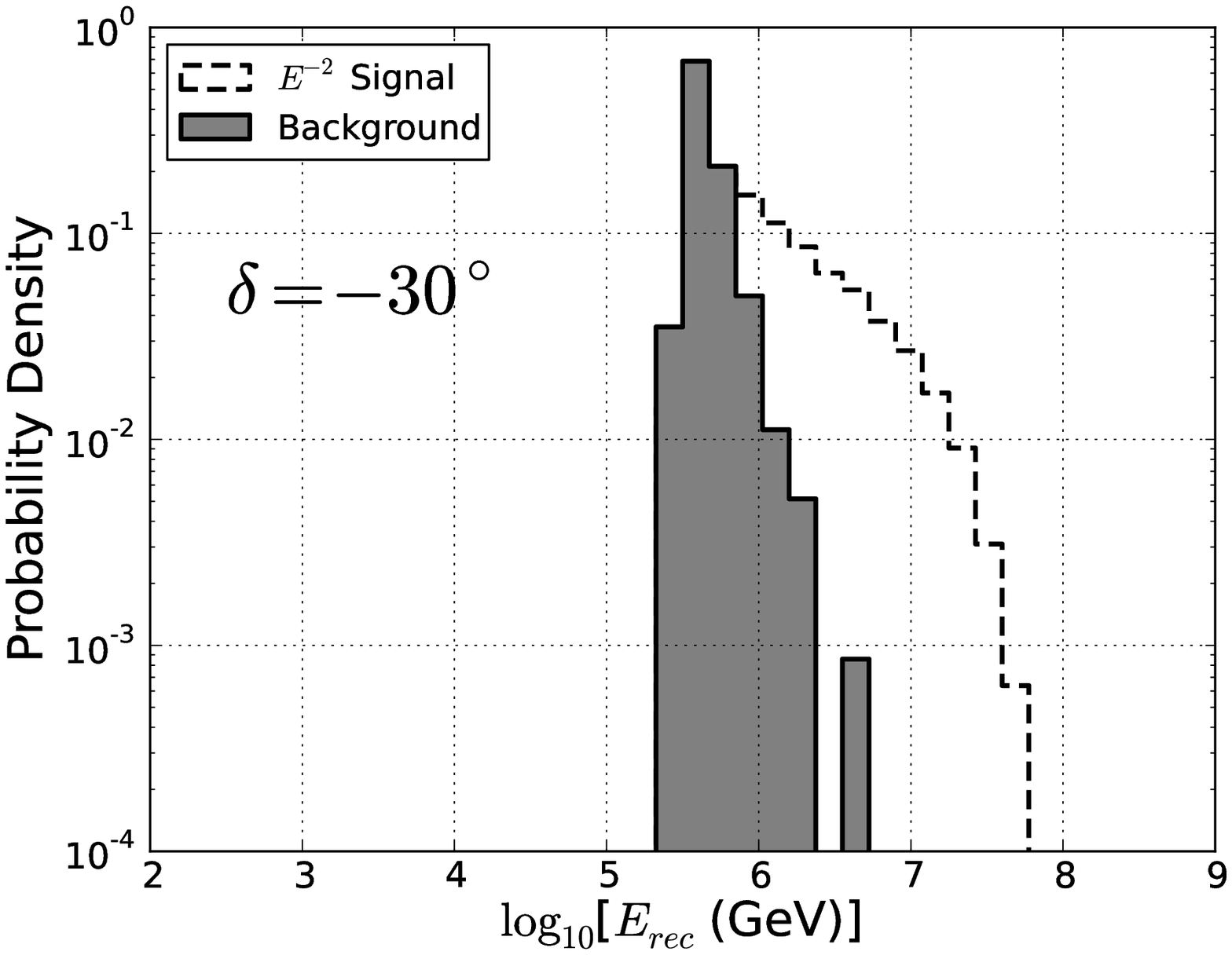} 
  \includegraphics[width=.5\textwidth]{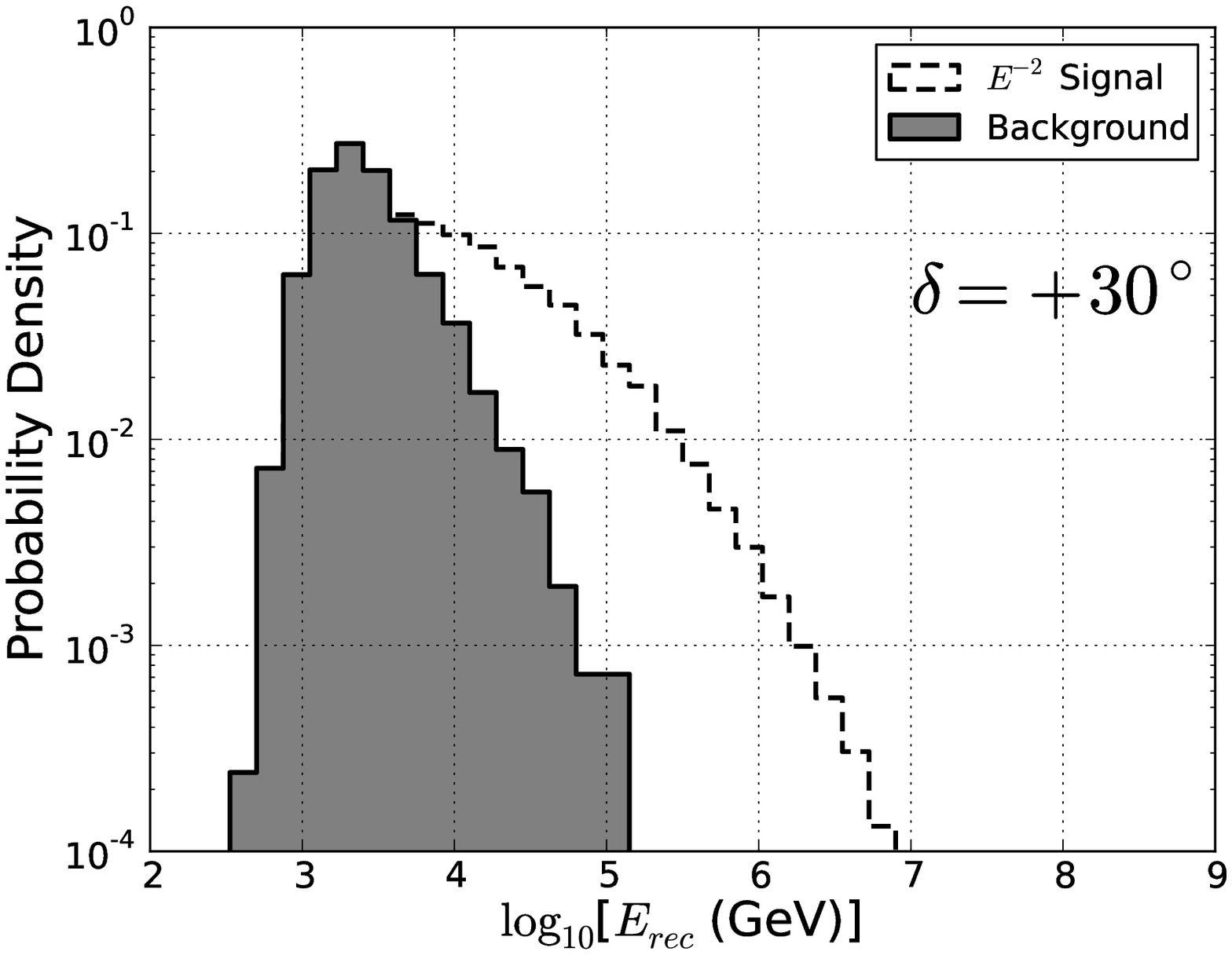} 
\end{tabular}
  \caption{Energy p.d.f. given as ${\rm d}N/{\rm d}\log_{10}E_{rec}$ for two different declinations,  $\delta=-30^{\circ}$ (left column) and $\delta=30^{\circ}$ (right column) for background and an exemplary signal of an E$^{-2}$ spectrum for the three different detector configurations, the 79-string configuration (top row), the 59-string configuration (middle row) and the 40-string configuration (bottom row).}
  \label{fig:Eprob}
 \end{figure}

The energy p.d.f., $\mathcal{E}^{j}_{i}$, in the case of signal, is a
function of the reconstructed energy proxy, $E_{i}$, and the spectral index,
$\gamma$, of a power-law spectrum for a given declination (see
Fig.~\ref{fig:Eprob}).

 The background p.d.f., $\mathcal{B}^j_i$, is obtained from the experimental data and is given by:

\begin{equation}
\label{eq:backgroundpdf}
\mathcal{B}^j_i = B^{j}_{i} (\delta_{i}) \mathcal{E}^{j}_{i} (E_i, \delta_i).
\end{equation}

The space term, $B^{j}_{i} (\delta_{i})$, is the event density per unit solid angle as a function of the declination. The background density is right ascension independent due to the
Earth's rotation. The energy p.d.f. for background, $\mathcal{E}^{j}_{i}$, represents the probability of obtaining an energy $E_i$ from atmospheric backgrounds (neutrinos and muons) and therefore depends only on the declination.

The signal is considered to have the same spectrum for all data sets
and therefore the spectral index meets the condition of $\gamma_{j} = \gamma$. The fitted numbers of signal events $n^{j}_s$ in each sample are also fixed relative to each other, according to the signal hypothesis tested and the resulting fraction, $f^{j} (\gamma, \delta)$, of total signal events expected in each sample. Simulation is used to calculate this fraction of signal events coming from each data set for a given a spectral index, so that $n^{j}_s = f^{j}n_{s}$ (see Fig.~\ref{fig:fraction}). In this way, the likelihood, $\mathcal{L}$, remains a function of only the global parameters $n_s$ and $\gamma$ with respect to which it is maximized:

\begin{figure}[!th]
  \vspace{5mm}
  \centering
  \includegraphics[width=4.in]{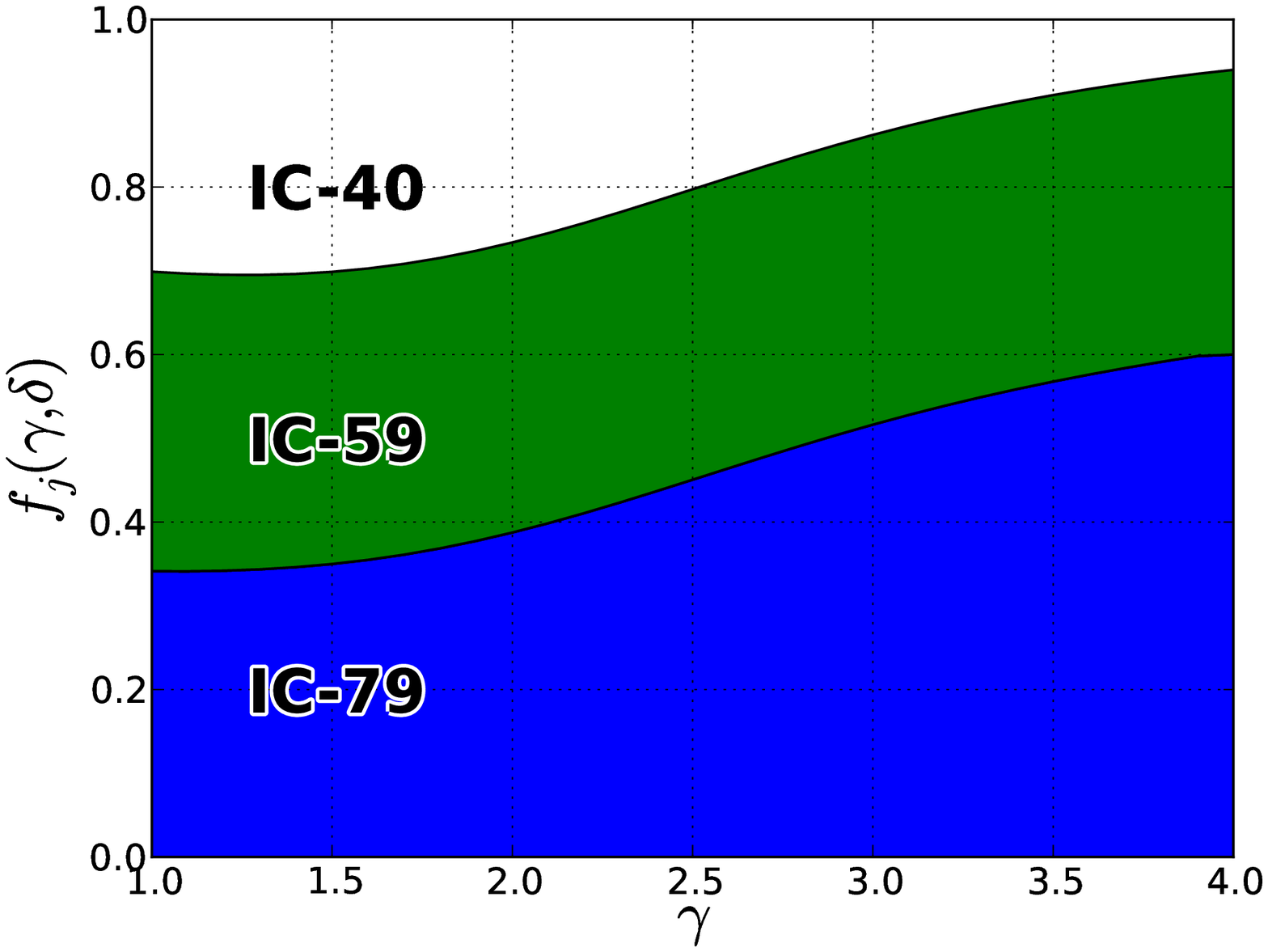}
  \caption{Relative efficiency or fraction of signal events coming from a source located at a declination of +16$^{\circ}$ as a function of the spectral index for the three configurations.  This relative efficiency is used as a weight in the likelihood method when combining multiple event selections.}
  \label{fig:fraction}
 \end{figure}

\begin{equation}
\label{eq:likelihood}
\mathcal{L}(\gamma, n_{s}) = \prod_{j} \mathcal{L}^{j} (\gamma, n^{j}_{s}) = \prod_{j}\;\prod_{i \in j} \left[\frac{n^{j}_{s}}{N^{j}}\mathcal{S}^{j}_{i} + \left(1- \frac{n^{j}_{s}}{N^{j}}\right)\mathcal{B}^{j}_{i} \right],
\end{equation}

where $i \in j$ indicates that the $i^{th}$-event is in sample $j$.
The test-statistic, $TS$, is calculated from the likelihood ratio of the background-only (null) hypothesis over the best fitted signal-plus-background hypothesis: 

\begin{equation}
\label{eq:TS}
TS = -2 \log \left[\frac{\mathcal{L}(n_s = 0)}{\mathcal{L}(\hat{n}_s, \hat{\gamma})}\right].
\end{equation}

Here, $\hat{n}_s$ is the best fit number of source events, and $\hat{\gamma}$ is the best fit spectral index. In principle, $n_s$ may be positive or negative since both positive and negative fluctuations with respect to the background expectation may be observed. In the likelihood maximization however, it is constrained to non-negative values. Pseudo-experiments on randomized data are performed to determine the significance of the observation. The randomization is achieved by the assignment of a random right ascension to each event in the data sample while all other event properties such as the energy and declination are left unchanged. The fraction of the pseudo-experiments which yield a $TS$ value above the observed $TS$ is quoted as the $p$-value of the observation.

For the stacking searches we used the method described in
Refs.~\cite{Stacking, IC40}. The signal p.d.f. is modified by breaking it
into a sum over $M$ sources. For one single sample the p.d.f can be re-written
as:

\begin{equation}
\label{eq:Stacking}
\mathcal{S}_i\rightarrow
\mathcal{S}^{tot}_i=\frac{\sum_{k=1}^MW^kR^k(\gamma,\delta_{k})\mathcal{S}^k_i(|\vec{x}_i-\vec{x}_{k}|,\sigma_{i}) \mathcal{E}_{i} (E_i, \delta_i, \gamma)}{\sum_{k=1}^MW^kR^k(\gamma,\delta_{k})},
\end{equation}

\noindent where $W^{k}$ is the relative theoretical weight for the $k$-th
source in the catalog and $R^{k}(\gamma,\delta_k)$ is the detector acceptance
for a flux with spectral index $\gamma$ at the coordinates
$\vec{x}_{k}$. The theoretical weights are chosen to minimize the flux required for discovery for a possible signal hypothesis. In catalogs where the predicted neutrino luminosity is strongly correlated with gamma-rays/X-ray/infrared
fluxes we use these observations as a base for the theoretical weights. For
catalogs with different possible theoretical flux predictions, the sources
can be weighted equally to maintain our sensitivity towards various signal
hypothesis. The spectral index, $\gamma$, is assumed to be the same for all
sources within a specific stacking search, and is a fit parameter along with
the total number of signal events $n_{s}$.

The following is a description of all the searches performed with the three
years of IceCube data (similar to those performed in Ref.~\cite{IC40}):
\begin{description}
\item[All-sky scan search.] An all-sky search, where the likelihood is
  evaluated in each direction in the sky in steps of $0.1^{\circ} \times
  0.1^{\circ}$ centered at the position of the source $\vec{x}_s$ over the
  declination range -85$^{\circ}$ to +85$^{\circ}$. In this search the number
  of effective trials is very high and related to the number of positions
  in the grid. The significance of an excess found in some direction needs to
  be corrected for these trials.
\item[A list of 44 selected sources.] In order to reduce the large number of
  effective trials associated with scanning the entire sky, we also performed
  a search for the most significant of 44 {\it a priori} selected source
  candidates. This source list is selected according to observations in gamma-
  rays or astrophysical modeling predicting neutrino emission.
\item[Stacking of 6 Milagro TeV gamma-ray sources.]  This catalogue is
  composed of most of the Milagro sources from \cite{milagro2007} 
  considered by the authors of Ref.~\cite{halzen2008} who estimated their neutrino emission.
  Given the observation in the IC-40 analysis of a significant {\it a
    posteriori} $p$-value from this catalogue, we considered a prescription for
  future samples and therefore the IC-40 data are not used in this
  analysis to avoid bias. Recent publications by the Milagro collaboration \cite{milagro2012} ruled out some of the
  assumptions about gamma-ray fluxes used in Ref.~\cite{halzen2008} so we use an equal weight for each source in the likelihood, with the
  intention of keeping our sensitivity optimal for all possible signal
  hypothesis.
\item[Stacking search for 127 local starburst galaxies.] This search was
  already performed using IC-40 data~\cite{IC40}. Starburst galaxies are
  interesting as possible neutrino sources due to their high star formation
  rates, especially of high mass stars. The large amount of stars leads to lots of SNRs, possibly the sites of CRs acceleration below the knee. In~\cite{Starburst} the authors associate the Far Infrared
  (FIR) emission with this hot ambient dust and the radio emission with
  synchrotron losses of electrons, which are assumed to be accelerated along
  with CRs in the large number of SNRs. The high star formation rate is
  believed to be the underlying cause for the observed strong correlation
  between the FIR and the radio flux, and hence the neutrino fluxes are
  expected to follow a similar pattern. We perform a stacking search for the
  catalog of 127 starburst galaxies as compiled in Table A.1
  in~\cite{Starburst}. We use the FIR flux at 60 $\mu m$ as the theoretical
  weight in the search hypothesis.
\item[Stacking search for 5 nearby clusters of galaxies] The stacking search
  for nearby clusters of galaxies, updated here after first results were
  presented in Ref.~\cite{IC40}, is performed by testing four models assuming
  different CR spatial distribution within the
  source~\cite{galaxycluster}. Clusters of galaxies are interesting potential
  sources of neutrinos that could be produced by interactions between high
  energy protons and the Intra Cluster Medium (ICM). In~\cite{galaxycluster}
  the authors discuss four different spectral shapes for the possible
  neutrino emission from these sources, as characterized by four different
  models of CR distribution. The source extensions are different for each
  model for different sources and are modeled as 2 dimensional Gaussian
  distributions with the corresponding widths for each model.  The
  differential fluxes predicted by~\cite{galaxycluster} are parametrized as
  broken power laws as described in~\cite{IC40} and used as theoretical
  weights in the likelihood.
\item[Stacking search for SNRs associated with molecular clouds.]  Molecular
  clouds surrounding SNRs can serve as target for high energy protons (or heavier nuclei) accelerated by SNR shocks to produce high energy gamma-rays and neutrinos. Specific models
  such as~\cite{MC} suggest a high correlation between the expected gamma-ray
  and neutrino fluxes. We stack sources from a catalog of close molecular
  clouds associated with SNRs, which were observed at high energy by AGILE,
  Fermi, VERITAS, H.E.S.S. and MAGIC~\cite{W51CFermi, W51CMilagro, W51CHess, W44Fermi}. 
  The expected neutrino energies from these sources do not allow for a study in the Southern Hemisphere where this search has
  sensitivity to PeV - EeV energies. Only galactic sources in the northern sky, where IceCube is sensitive to TeV
  energies, were selected. The catalog contains 4 SNR associated with molecular clouds:
  W51C, W44, IC 443 and W49B. The integrated gamma-ray flux above 1 TeV for
  each source (in Crab units) is used as the theoretical weight in the
  likelihood. Very recently the Fermi collaboration detected the
  characteristic pion-decay signature in the gamma-ray emission for two of
  these SNRs, IC 443 and W44, providing direct evidence that Cosmic Rays
  protons are accelerated in SNRs~\cite{FermiSNR} at GeV energies.
\item[Stacking search for galaxies with supermassive black holes.] Ref.~\cite{bh}
  systematically catalogs possible black hole candidates from within the
  Greisen-Zatsepin-Kuzmin limit \cite{GZK} of around 100 Mpc. In order
  to keep only the most powerful emitters of particles, a cut of
  $5\times10^8$ solar masses is applied to the catalog to produce a final
  list of 233 sources. We use as weights the 2 micron Near
  Infrared flux from the 2 Micron All Sky Survey that can be
  related to the mass of black holes~\cite{bh}.

\end{description}

\section{Results}
\label{sec5}

\begin{figure}[t!]
\hspace{-2cm}  
  \includegraphics[width=1.25\textwidth]{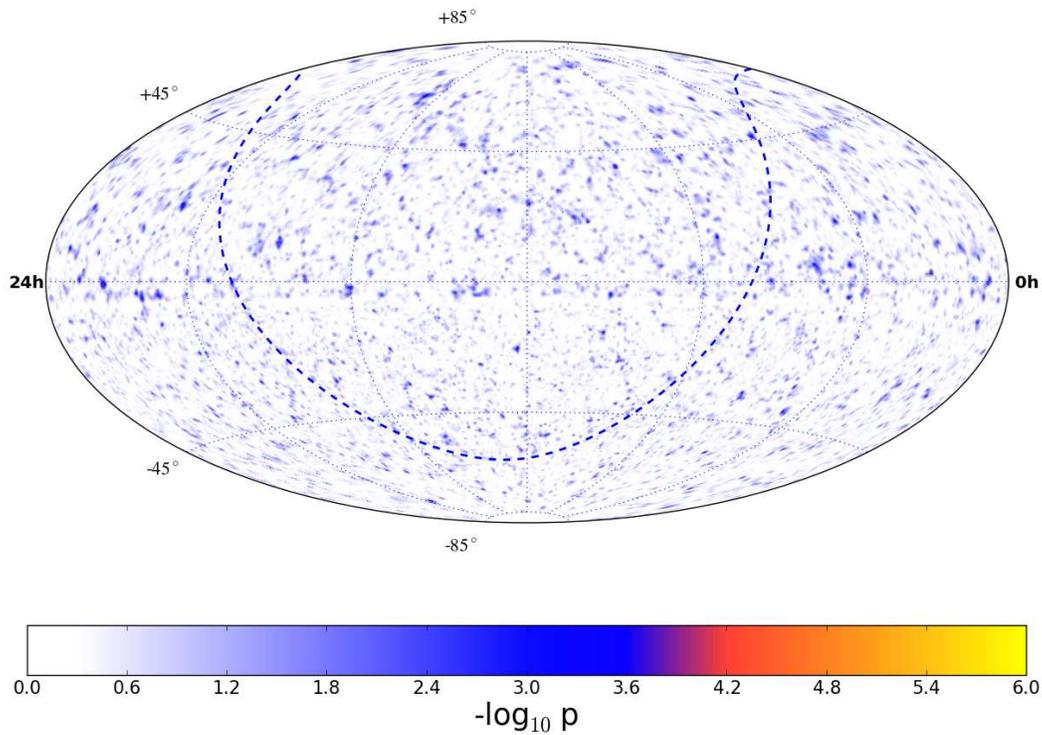}
\vspace{-1cm}
  \caption{Pre-trial significance skymap in equatorial coordinates (J2000) of the all-sky point source scan for the combined IC79+IC59+IC40 data sample. The dashed line indicates the galactic plane.}
  \label{fig:skymap}
\end{figure}

The results of the all-sky scan are shown in the pre-trial significance map of $p$-values in Fig.~\ref{fig:skymap}.
The most significant deviation in the northern sky has a pre-trial $p$-value of 1.96~$\times$~10$^{-5}$ and is located at 34.25$^{\circ}$ r.a. and $2.75^{\circ}$ dec. Similarly, the most significant deviation in the southern sky has a pre-trial $p$-value of 8.97~$\times$~10$^{-5}$ and is located at 219.25$^{\circ}$ r.a. and $-38.75^{\circ}$ dec.

The post-trial probabilities calculated as the fraction of scrambled sky maps with at least one spot with an equal or higher significance for each hemisphere correspond to 57\% and 98\% for the northern and the southern spots respectively and therefore both excesses are well compatible with the background hypothesis. 
\begin{figure}[t!]
\begin{tabular}{c c}
\hspace{-0.5cm}
  \includegraphics[width=0.54\textwidth]{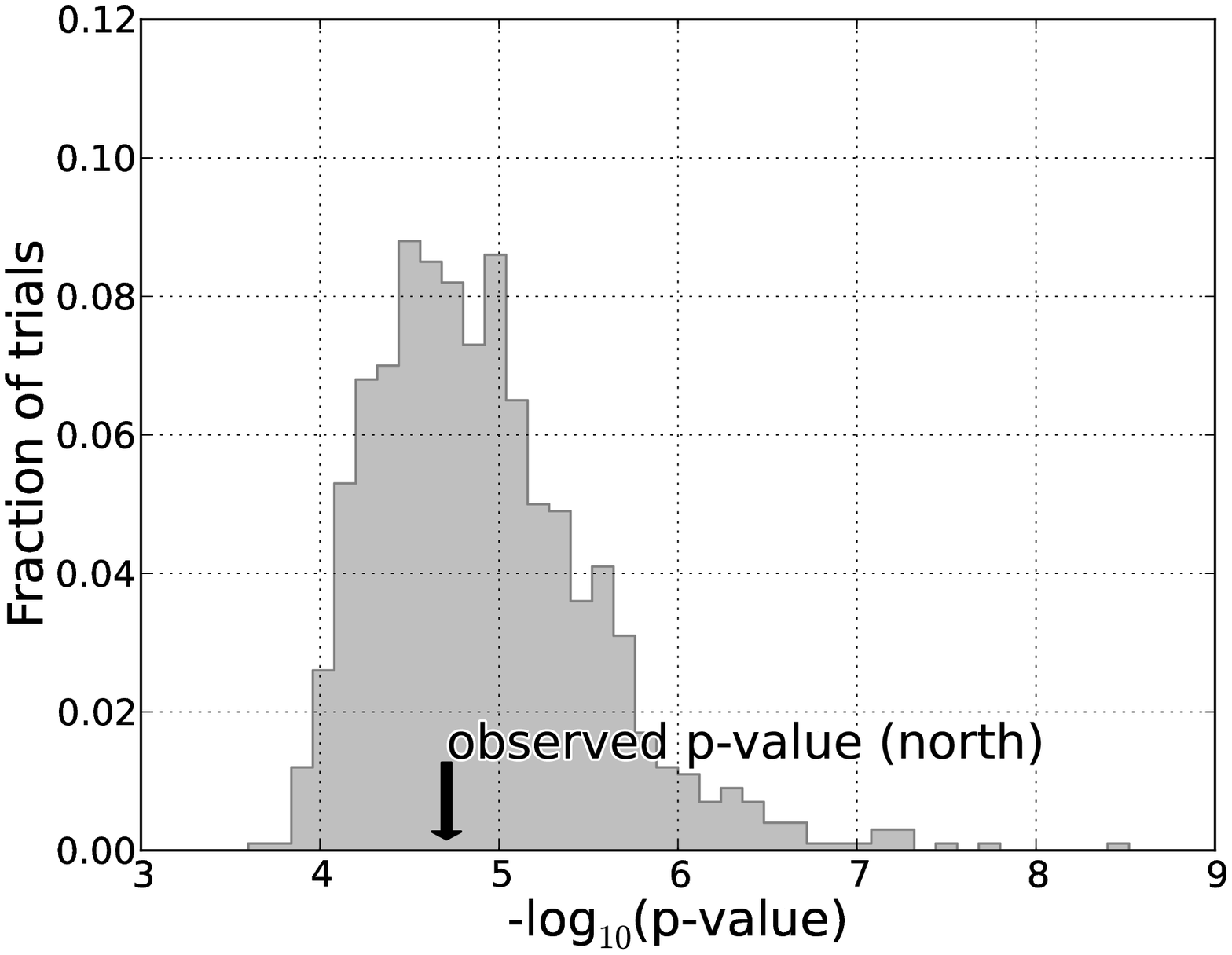}
\hspace{-0.5cm}
    \includegraphics[width=0.54\textwidth]{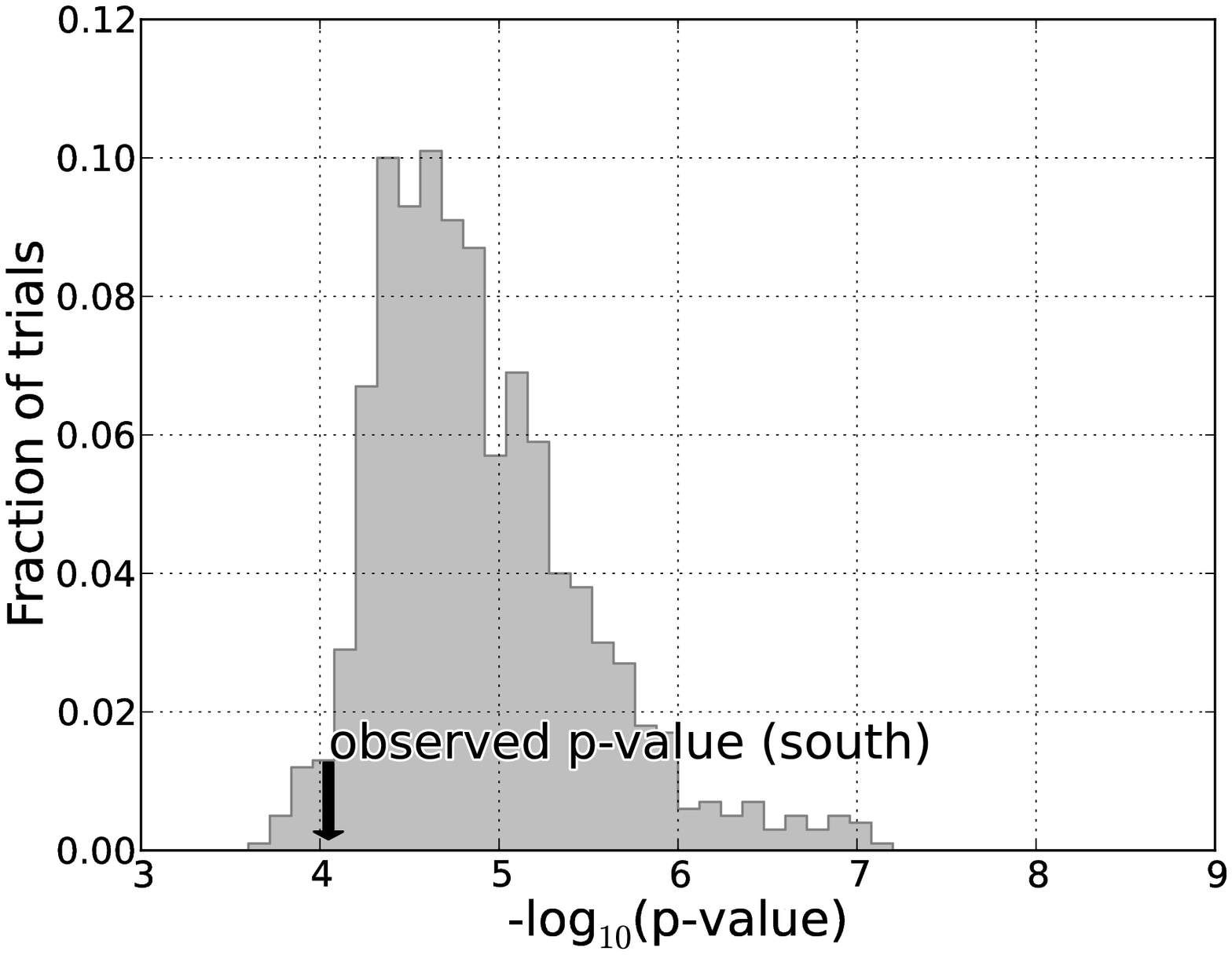}
\end{tabular}
\caption{The distribution of the smallest $p$-value in the Northern (left) and the Southern (right) Hemisphere, obtained from randomized data. The observed $p$-values of the two hottest spots in the data are indicated by the two arrows.}
\label{fig:posttrials}
\end{figure}

Figure~\ref{fig:posttrials} shows the $p$-value distribution for the hottest spot in the Northern Hemisphere (left) and for the Southern Hemisphere (right). The observed $p$-value in the data is indicated in both distributions, the final post-trial is given by integrating the right hand side of the distribution from the observed values.

\begin{figure*}[t!]
  \centering
  \includegraphics[width=5in,height=4in]{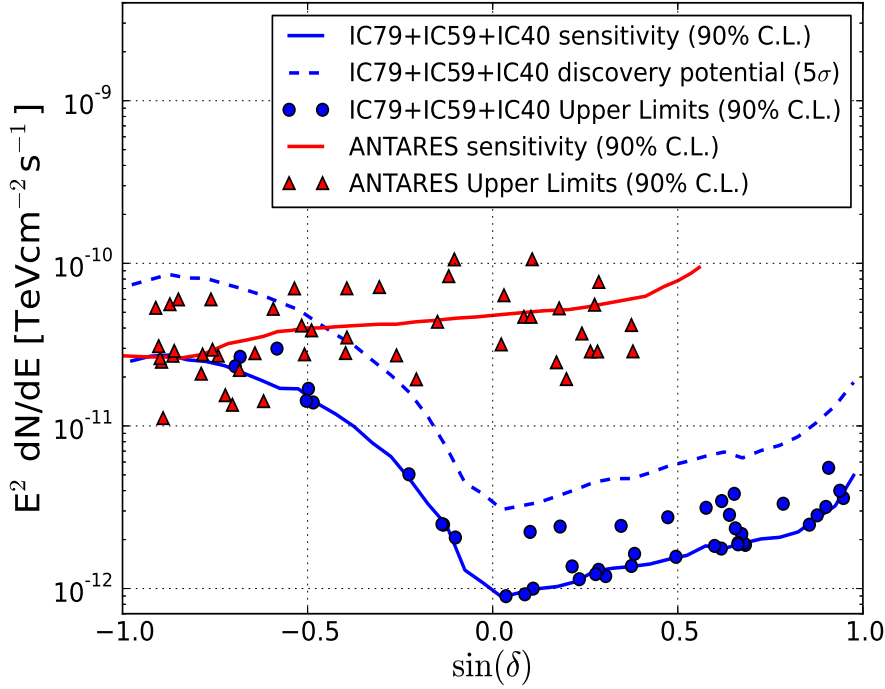}
  \caption{Muon neutrino and antineutrino flux 90\% C.L. upper limits and sensitivities for an E$^{-2}$ spectrum. Published limits of ANTARES~\cite{ANTARES} are shown. The different likelihood function and method to derive upper limits used by ANTARES may account for differences in the limits from the two experiments at the level of 20\%. In the case of the IceCube method, negative values of the number of signal events are not allowed in the minimization procedure. Therefore for those sources where there was an under-fluctuation of the background the upper limit matches the median upper limit.}
  \label{fig:ul}
\end{figure*}

The results of the point-source search in the direction of the 44 search selected {\it a priori} according to the positions of known objects is summarized in Tab.~\ref{tab:gul} and ~\ref{tab:eul}. The smallest $p$-value in the northern sky is found in the direction of HESS J0632+057 with a probability of 5.8\%, however this value is translated into a post-trial probability of 65\% once it is compared with an ensemble of randomized sky maps. For the southern sky, the highest significance is observed at the position of PKS 1454-354 with a pre-trial $p$-value of 23\% which corresponds to a post-trial probability of 70\%. The forth column of Tab.~\ref{tab:gul} and \ref{tab:eul} shows the upper limits for an E$^{-2}$ flux of $\nu_{\mu} + \bar{\nu}_{\mu}$ calculated at 90\% C.L. based on the classical (frequentist) approach~\cite{Neyman} for each of the selected objects. The same values are indicated in Fig.~\ref{fig:ul} together with the IceCube sensitivity defined as the median upper limit and the discovery potential. Also shown are the ANTARES upper limits for a list of locations~\cite{ANTARES}

\begin{center}
\vspace{0.5cm}
\begin{longtable}[t!]{ p{1.7cm}  | l |r | r | c | c | c |c | c  }
\caption{\label{tab:gul} Results for galactic objects on the {\it a priori} search list. }\\
\toprule
Category & Source & r.a. [$^{\circ}$] & dec. [$^{\circ}$] & $p$-value & $\hat{n}_{S}$ & $\hat{\gamma}$  & B$_{1^{\circ}}$ & $\Phi_{\nu_{\mu} + \bar{\nu}_{\mu}}^{90\%}$ \\
\midrule[\heavyrulewidth]
\endfirsthead
\multicolumn{9}{c}%
{\tablename\ \thetable\ -- \textit{Continued from previous page}} \\[{0.5cm}]
\toprule
Category & Source & r.a. [$^{\circ}$] & dec. [$^{\circ}$] & $p$-value & $\hat{n}_{S}$ & $\hat{\gamma}$ & B$_{1^{\circ}}$ & $\Phi_{\nu_{\mu} + \bar{\nu}_{\mu}}^{90\%}$ \\
\midrule[\heavyrulewidth]
\endhead
\bottomrule \multicolumn{9}{r}{\textit{Continued on next page}} \\[{0.5cm}]
\endfoot
\hline
\multicolumn{9}{l}{}\\[{0.1cm}]
\multicolumn{9}{l}{\parbox{\textwidth}{\textit{ Note. -- } Sources are grouped according to their classification as High-Mass X-ray binaries or micro-quasars (HMXB/mqso),  SNRs, Pulsar Wind Nebulas (PWNs), star formation regions and unidentified sources. The $p$-value is the pre-trial probability of compatibility with the background-only hypothesis. The $\hat{n}_{S}$ and $\hat{\gamma}$ columns give the best-fit number of signal events and spectral index of a power-law spectrum. When $\hat{n}_{S} = 0$ no $p$-value or $\hat{\gamma}$ are reported.
The eighth column gives the number of background events in a circle of 1$^{\circ}$ around the search coordinates. The last column shows the upper limits based on the classical approach \cite{Neyman} for an E$^{-2}$ flux normalization of $\nu_\mu + \bar{\nu}_{\mu}$ flux in units of $10^{-12}$ TeV$^{-1}$ cm$^{-2}$s$^{-1}$.\\
    }}\\[{1cm}]
\multicolumn{9}{l}{\parbox{\textwidth}{$^{a}$Most significant $p$-value in the northern sky among all galactic and extragalactic objects on the {\it a priori} search list.}} \\

\endlastfoot
       
   {SNR} &  TYCHO &   6.36 & 64.18  & -- & 0.0 & -- & 11.1& 3.18\\
      &         Cas A & 350.85 & 58.81 & -- & 0.0 & -- & 11.5& 2.47 \\
      &         IC443 &  94.18 & 22.53  & 0.43 & 2.8 & 3.9 & 17.2& 1.63\\
           \midrule         
  {HMXB} 
&  LSI +63 303 &  40.13 & 61.23  & -- & 0.0 & -- & 11.5& 2.82\\
/mqso &     Cyg X-3 & 308.11 & 40.96 & 0.43 & 2.5 & 3.9 & 12.9& 2.35 \\
   &          Cyg X-1 & 299.59 & 35.20  & 0.21 & 5.6 & 3.9 & 14.6& 3.14\\
  & HESS J0632+057 &  98.25 &  5.80 & 0.058$^{a}$ & 15.6 & 3.4 & 24.1& 2.23 \\
 &              SS433 & 287.96 &  4.98 & -- & 0.0 & -- & 24.3& 0.92 \\
      \midrule
     {Star Formation Region} 
     &   Cyg OB2 & 308.08 & 41.51  & -- & 0.0 & -- & 12.7& 1.87\\
\midrule         
  {pulsar/} 
  &  MGRO J2019+37 & 305.22 & 36.83  & -- & 0.0 & --& 14.3& 1.83\\
{PWN} & Crab Nebula &  83.63 & 22.01  & -- & 0.0 & --& 17.2& 1.38\\
\newpage
&  Geminga &  98.48 & 17.77  & -- & 0.0 & -- & 19.5& 1.193\\
 \midrule
    {Galactic Center}
       &        Sgr A* & 266.42 & -29.01 & 0.49 & 0.6 & 3.7 & 25.2& 13.94 \\
  \midrule
   {Not identified} & MGRO J1908+06 & 286.98 &  6.27 & -- & 0.0 & -- & 23.8& 1.00 \\

\end{longtable}
\end{center}

\begin{center}
\vspace{-1cm}
\begin{longtable}[th!]{ p{1.7cm}  | l |r | r | c | c | c |c | c  }
\caption{\label{tab:eul} Results for extragalactic objects on the {\it a priori} search list. }\\
\toprule
Category & Source & r.a. [$^{\circ}$] & dec. [$^{\circ}$] & $p$-value & $\hat{n}_{S}$ & $\hat{\gamma}$  & B$_{1^{\circ}}$ & $\Phi_{\nu_{\mu} + \bar{\nu}_{\mu}}^{90\%}$ \\
\midrule[\heavyrulewidth]
\endfirsthead
\multicolumn{9}{c}%
{\tablename\ \thetable\ -- \textit{Continued from previous page}} \\[{0.5cm}]
\toprule
Category & Source & r.a. [$^{\circ}$] & dec. [$^{\circ}$] & $p$-value & $\hat{n}_{S}$ & $\hat{\gamma}$ & B$_{1^{\circ}}$ & $\Phi_{\nu_{\mu} + \bar{\nu}_{\mu}}^{90\%}$ \\
\midrule[\heavyrulewidth]
\endhead
\bottomrule \multicolumn{9}{r}{\textit{Continued on next page}} \\[{0.5cm}]
\endfoot
\hline
\multicolumn{9}{l}{}\\[{0.1cm}]
\multicolumn{9}{l}{\parbox{\textwidth}{\textit{ Note. -- }  Sources are grouped according to their classification as BL Lac objects, Radio Galaxies, Flat-Spectrum Radio Quasars (FSRQ) and Starburst galaxies. The $p$-value is the pre-trial probability of compatibility with the background-only hypothesis. The $\hat{n}_{S}$ and $\hat{\gamma}$ columns give the best-fit number of signal events and spectral index of a power-law spectrum. When $\hat{n}_{S} = 0$ no $p$-value or $\hat{\gamma}$ are reported.
The eighth column gives the number of background events in a circle of 1$^{\circ}$ around the search coordinates. The last column shows the upper limits based on the classical approach \cite{Neyman} for an E$^{-2}$ flux normalization of $\nu_\mu + \bar{\nu}_{\mu}$ flux in units of $10^{-12}$ TeV$^{-1}$ cm$^{-2}$s$^{-1}$.\\
    }}\\[{1cm}]
\multicolumn{9}{l}{\parbox{\textwidth}{$^{b}$Most significant $p$-value in the southern sky among all galactic and extragalactic objects on the {\it a priori} search list.}}\\

\endlastfoot
{BL Lac} & S5 0716+71 & 110.47 & 71.34 & -- & 0.0 & -- & 10.3& 3.60 \\
	      & 1ES 1959+650 & 300.00 & 65.15 & 0.19 & 5.7 & 3.9 & 11.1& 5.53 \\
               &  1ES 2344+514 & 356.77 & 51.70 & 0.29 & 4.7 & 3.9 & 12.4& 3.32 \\
&  3C66A &  35.67 & 43.04  & -- & 0.0 & -- & 12.7& 1.86\\
      &  H 1426+428 & 217.14 & 42.67  & -- & 0.0 & -- & 12.7& 1.90\\
      &   BL Lac & 330.68 & 42.28 & 0.42 & 3.7 & 3.3 & 12.7& 2.16 \\
       &  Mrk 501 & 253.47 & 39.76  & 0.34 & 4.8 & 3.9 & 13.4& 2.84\\
       &  Mrk 421 & 166.11 & 38.21  & 0.18 & 3.7 & 1.8 & 13.7& 3.45\\
       & W Comae & 185.38 & 28.23  & 0.21 & 2.8 & 1.8 & 16.1& 2.74\\
       & 1ES 0229+200 &  38.20 & 20.29 & 0.19 & 8.2 & 3.9 & 17.8& 2.43 \\
       & PKS 0235+164 &  39.66 & 16.62  & -- & 0.0 & -- & 19.9& 1.30\\
       &  PKS 2155-304 & 329.72 & -30.23  & -- & 0.0 & -- & 25.5& 14.28\\
       &   PKS 0537-441 &  84.71 & -44.09 & -- & 0.0 & -- & 23.8 & 23.27\\
\midrule
{FSRQ} 
&             4C 38.41 & 248.81 & 38.13  & -- & 0.0 & -- & 13.7& 1.76\\
&         3C 454.3 & 343.49 & 16.15 & -- & 0.0 & -- & 19.9& 1.23 \\
   &     PKS 0528+134 &  82.73 & 13.53 & -- & 0.0 & -- & 20.8& 1.14 \\
 &         PKS 1502+106 & 226.10 & 10.49  & 0.076 & 8.4 & 2.3 & 21.2& 2.40\\
 &             3C 273 & 187.28 &  2.05  & -- & 0.0 & -- & 25.0& 0.90\\
  &             3C279 & 194.05 & -5.79 & -- & 0.0 & -- & 23.5& 2.06 \\
   &     QSO 2022-077 & 306.42 & -7.64 & -- & 0.0 & -- & 23.2& 2.47 \\
     &   PKS 1406-076 & 212.24 & -7.87  & -- & 0.0 & -- & 23.2& 2.49\\
    &    QSO 1730-130 & 263.26 & -13.08& -- & 0.0 & -- & 25.6 & 5.04 \\
     & PKS 1622-297 & 246.53 & -29.86  & 0.45 & 0.7 & 4.0 & 25.2& 16.91\\
     &    PKS 1454-354 & 224.36 & -35.65& 0.23$^{b}$ & 1.0 & 5.9 & 24.1 & 29.89 \\
 \midrule
    {Starburst}   &    M82 & 148.97 & 69.68 & -- & 0.0 & -- & 10.7& 4.00 \\ 

\midrule
     {Radio}
  & NGC 1275 &  49.95 & 41.51 & -- & 0.0 & -- & 12.7& 1.91 \\
Galaxies &     Cyg A & 299.87 & 40.73 & 0.15 & 1.5 & 1.5 & 12.9& 3.82 \\
  &             Cen A & 201.37 & -43.02 & 0.46 & 2.0 & 1.4 & 23.9 & 26.62\\
         \newpage
       &    3C 123.0 &  69.27 & 29.67  & -- & 0.0 & -- & 15.9& 1.57\\
       &  M87 & 187.71 & 12.39 & 0.45 & 2.9 & -- & 20.9& 1.37 \\
      \end{longtable}
\end{center}

The 6 Milagro TeV sources stacking analysis resulted in a post-trial $p$-value of 20.4\% with a best fit $\hat{n}_{s} = 17$. In the GC stacking searches, less events than expected from the background were observed for all of the four models tested, meaning that the $p$-value is at least $\ge 50$\%. Also the SNRs associated with Molecular Clouds as well as the Starburst galaxies resulted both in negative fluctuations of the background in every case with $\hat{n}_s = 0$. Finally the black hole stacking search produced a post-trial $p$-value of 44.3\% with 12 signal events as the best fit.


\subsection{Implications for models of astrophysical neutrinos.}

\begin{figure*}[t!]
  \centering
  \includegraphics[width=.7\textwidth]{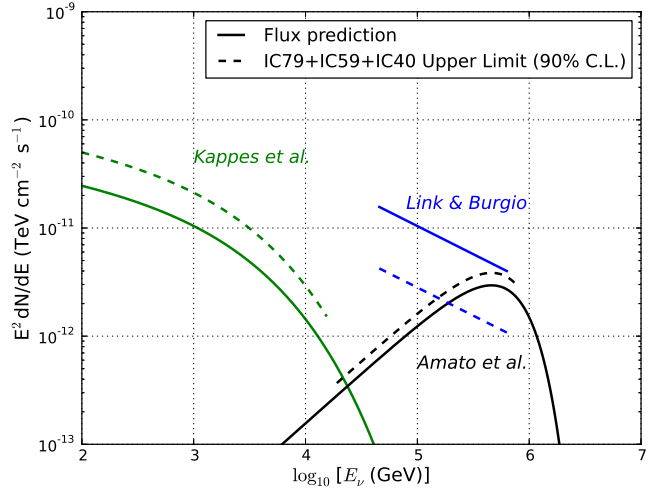}
  \caption{Predicted muon neutrino fluxes for several hadronic models for steady neutrino emission from the Crab and upper limits based on 3 years of IceCube data. Solid lines indicates the flux prediction and the dashed lines the corresponding upper limit flux for a 90\% C.L. for an energy range that contains 90\% of the signal. Neutrino oscillations are accounted for.}
  \label{fig:crabmodels}
\end{figure*}

This analysis has shown that there is
no evidence of neutrino emission from point-sources in the sky. In the
absence of a positive detection it is, however, possible to constrain some
models that predict astrophysical neutrino emissions. IceCube is entering a
new stage in which a non-discovery has meaningful implications and can
provide insight about the nature of these phenomena. IceCube has provided the
most constraining upper limits on neutrino fluxes from sources like the Crab
~\cite{Crab}. Even though the Crab spectral emission seems to be fully
explained by electromagnetic phenomena, several $\gamma$-ray flares observed
in the past years in the GeV region ($E_{\gamma}> 100$~MeV) challenge purely
leptonic models \cite{Fermi_flare}. The impact of IceCube limits on different
models of neutrino emission from the Crab was already discussed
in~\cite{Crab} for the 40-string configuration of IceCube. Here we update the
upper limits based on this three year analysis of
IceCube. Figure~\ref{fig:crabmodels} summarizes a number of different
predicted muon neutrino fluxes at Earth according to several models (standard
oscillations have been taken into account). The green solid line corresponds
to the flux predicted in~\cite{Crabkappes} based on the $\gamma$-ray spectrum
measured by H.E.S.S. As can be seen, the IceCube upper limit is only a factor
of two above the flux prediction. This is interesting since it indicates
that neutrino astronomy is at the level of sensitivity of gamma astronomy
experiments (the factor of two corresponds to the muon neutrino flux lost due
to oscillations along the path from the source).  The black line represents
the estimated flux based on the resonant cyclotron absorption model. In
Ref.~\cite{CrabBlasi} inelastic nuclear collisions are
considered and the predicted neutrino rates depend on the Lorentz factor,
$\Gamma$, of nuclei injected by the pulsar and the effective target
density. The predicted flux in Fig.~\ref{fig:crabmodels} is for the most
optimistic case of the effective target density and a wind Lorentz factor of
$\Gamma = 10^{7}$.  Ref~\cite{CrabBurgio, Link:2006pd} considers scattering of wind protons with the X-ray emission from the pulsar's surface. The predicted neutrino flux assuming a quadratic scaling of the proton's energy with the height above the surface is shown in the plot. The most optimistic version of this model can be rejected with more than 90\% C.L.

\begin{figure*}[ht!]
 \centering
  \includegraphics[width=0.7\textwidth]{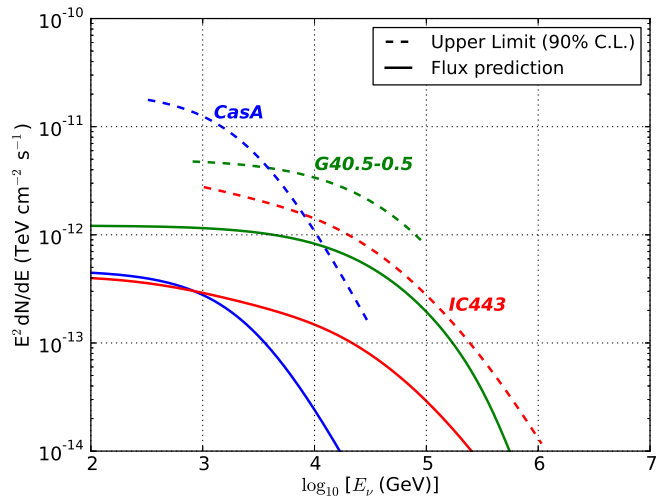}
  \caption{Predicted muon neutrino fluxes from three modeled SNRs in the Northern Hemisphere according to~\ref{fig:snrmodels}. The muon neutrino 90\% C.L. upper limits from this analysis are shown in the energy range of the 90\% signal containment.}
  \label{fig:snrmodels}
\end{figure*}

IceCube upper limits are approaching to some predictions from models on
neutrino emission from SNRs. In~\cite{Julia} the authors calculate the
neutrino spectra generated by proton-proton interactions at supernova
remnants. For the Northern Hemisphere, the G40.5-0.5 seems to be the most promising candidate for a
neutrino detection due to the high photon flux from this source. The electron
spectrum of this supernova remnant is supposed to cut off at energies lower
than the measured radiation, indicating a possible hadronic origin of the radiation.
Figure~\ref{fig:snrmodels} shows the different predicted
muon neutrino spectra after considering oscillations for three SNRs. The 90
\% C.L. flux upper limit for muon neutrinos is also shown.

\begin{figure}[ht]
\centering
  \includegraphics[width=0.7\textwidth]{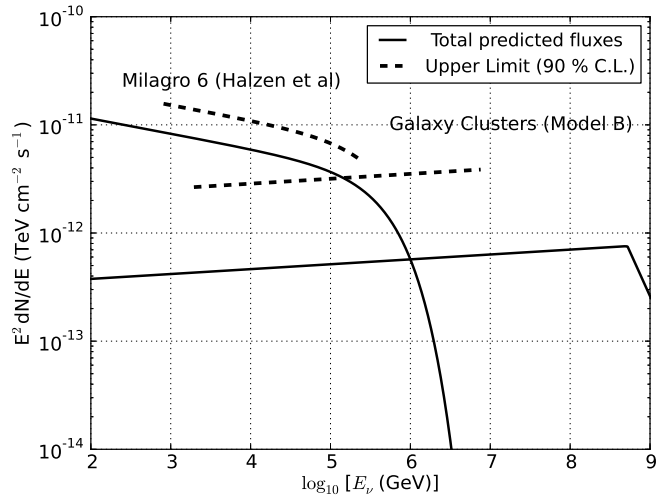}
\caption{Predicted muon neutrino fluxes from 6 Milagro sources in gamma-rays according to~\cite{halzen2008} and from the 5 nearby galaxy clusters considered in~\cite{galaxycluster}. The corresponding 90\% C.L. flux upper limit for muon neutrinos obtained from the stacking analysis are shown as well.}
  \label{fig:stacking}
\end{figure}

As can be seen, IceCube upper limits in the most optimistic case, for the
G40-5.0.5, are still a factor of four above the flux prediction. However,
stacking techniques can improve the discovery
potential. Figure~\ref{fig:stacking} shows the upper limit of the
stacking result of the 6 Milagro TeV gamma-ray associations assuming the
model from~\cite{halzen2008}. The result of the analysis was a positive
fluctuation, so the sensitivity is expected to be closer to the
total predicted flux than the upper limit. Together in this plot we show the flux prediction and the corresponding upper limit from the 5 nearby galaxy clusters search assuming that CRs are uniformly distributed within the virial radius of the galaxy cluster.

\section{Systematic Uncertainties}
\label{sec6}
One of the strengths of the presented searches is that they use a data-driven
background estimation based on randomized data. The p-values are unaffected
by uncertainties on the theoretical estimate of fluxes of the background of
atmospheric neutrinos and muons that depend on hadronic models of shower
development in the atmosphere and on the CR composition. They are also
unaffected by the poorly known contribution of prompt neutrinos. Moreover,
uncertainties on the simulation of the detector also do not affect the
post-trial p-value.

On the other hand upper limits are affected by the systematic errors on the
simulation of the detector efficiency and response to the flux of
neutrinos. The construction of the signal energy density function in the
likelihood method depends on simulation and is therefore affected by the
systematic uncertainties. In order to capture the impact of the systematic
uncertainties, we fully propagated each of them through the likelihood search
and calculated the sensitivity of the search for a discrete set of simulated
signal responses within the allowed range of uncertainties. We used the IC-79
data sample for this evaluation and we quote the declination-averaged
uncertainties under the assumption that all samples are affected by the same systematic errors.
This assumption can be regarded as conservative since the lower energy range is more strongly affected by 
the uncertainties and the IC-79 sample contains the largest number of low energy events.

The two most relevant uncertainties come from the absolute efficiency of the optical modules and the modeling of the optical properties of the ice. As a conservative estimate, we allowed for
a $\pm 10$\% uncertainty in the absolute efficiency of the optical
modules. Uncertainties in the relative sensitivity of the individual DOMs
with respect to the detector average have been observed to have a negligible
impact on the total flux uncertainty in the energy range of this
analysis. Likewise, there is no significant impact if the sensitivity of the
high quantum efficiency PMTs in DeepCore \cite{DeepCore} is larger with
respect to the rest of the detector. This is due to the limited size of this
part of the detector with respect to the typical track length of the event
selected in this analysis. The uncertainty of $\pm 10$\% in DOM
efficiency in simulation resulted in +6\%/-7\% variation in the sensitivity
of IC-79.
The parametrization of the optical properties of
the ice used in this work is a variant of the parametrization presented in
\cite{SPICE}. Its uncertainties have been taken to be $\pm 10$\% in
absorption and scattering and they both have been rescaled at the same time.
The effect in sensitivity coming from these variation was of +5\%/-8\%.

Due to constraints in computing power, we used tabulated photon arrival
probabilities in the signal simulation~\cite{photonics}. A more accurate
description of the detector response can be obtained by using simulation with
direct photon propagation~\cite{ppc}. The difference between the two is most
relevant for energies below $\sim$ 1 TeV and decreases with energy. In order
to quantify the impact of the photon propagation method, we compared the
difference in sensitivity in the northern sky using simulated data generated
specifically for this purpose. The impact on the Southern Hemisphere is
expected to be smaller and the values for the northern sky do thus represent
a conservative estimate for the full sky. The difference in sensitivity, 7.2\%,
between the two propagators can be accounted for by the uncertainty in the optical
efficiency and therefore here is not considered as an additional source of
systematic uncertainty. Future simulations of IceCube are expected to be produced with direct photon propagation while an increase in the nominal optical efficiency of 10\% is also foreseen since a higher optical efficiency was found to better describe IceCube data.

There is a small probability that southern sky signal neutrinos are vetoed by
the IceTop veto applied in the IC-79 and IC-59 data samples due to random
coincidences. As can be seen in Figure~\ref{fig:icetopveto_79} on the
left, this probability for random coincidences is constantly below 1\% at all
declinations and can therefore be neglected in comparison to the impact of
other systematic uncertainties.

By summing in quadrature all the different contribution the expected
uncertainty in the IC-79 sensitivity is about 18\%. This is compatible with
the 16\% estimated for the IC-40 configuration \cite{IC40}.

The upper limits listed in the previous section have been calculated for a pure muon neutrino
signal, under the assumption that no other neutrino flavors contribute in
this analysis. Considering neutrino oscillations with a large mixing angle
$\Theta_{23}\sim45^{\circ}$ and a long baseline, a typical neutrino flavor
ratio of $\nu_{e}$:$\nu_{\mu}$:$\nu_{\tau}$ = 1:2:0 at the source will result
in an approximate partition of 1:1:1 at Earth. In the case of $\nu_{\tau}$
the resulting $\tau$ will decay into a $\mu$ with a branching ratio of
about 17\%. These additional muons from $\nu_{\tau}$ can contribute to a
possible signal flux in this analysis. In \cite{IC40}, the contribution of
$\nu_{\tau}$ in addition to the $\nu_{\mu}$ flux simulated in this work has
been determined to be $10-16$\% of the $\nu_{\mu}$ contribution.

\section{Conclusions}
\label{sec7}

We present the results of the point source analysis of three years of data with the 40-string, 59-string and 79-string configurations of the IceCube Neutrino Observatory. The combined data has a total live-time of 1,040 days from April 2008 to May 2011. The all-sky survey found no evidence of point-source neutrino emission in the Northern or the Southern Hemisphere. The post-trial probabilities of the highest significant coordinate in each hemisphere are compatible with background fluctuations. Additionally, a search on a catalog of known emitters of high-energy radiation was performed. Several stacking analyses were carried out to integrate the possible signal from all sources of the same class. Also in this case, no significant deviation from the background hypothesis was found and the corresponding 90\% C.L. upper limits on the muon neutrino fluxes were calculated and compared to predictions. The most optimistic models considered here can be excluded at 90\% C.L. and in other cases limits are factor two to four above the predictions. 

The muon neutrino upper limits presented here improve earlier results~\cite{IC40} by a factor $\sim 3.5$ or better and are the strictest neutrino limits to date over the entire sky. Some of these limits for an E$^{-2}$ muon neutrino flux have reached the level of 10$^{-12}$ TeV cm$^{-2}$s$^{-1}$ necessary to test current models of neutrino emission expected for galactic sources like supernova remnants~\cite{SNR}.
In the future, the sensitivity of IceCube to neutrino point source will improve with the inclusion of additional data collected with the full IceCube array.

\appendix
\section{Muon neutrino effective area}
\label{AppAeff}

Table~\ref{tab:aeff} presents the tabulated values of the solid-averaged muon neutrino effective for the three different configurations used in this analysis: 

\begin{center}
\begin{longtable}[t] {c | c |r | r | r | r | r | r }
\caption{\label{tab:aeff} Muon neutrino effective areas.}
\\
\toprule
\multicolumn{2}{c}{ } & \multicolumn{3}{|c|}{North ($ 0^{\circ} < \delta \le 90^{\circ}$)}  & \multicolumn{3}{c}{ South ($ -90^{\circ} \le \delta \le 0^{\circ}$)} \\
\midrule
$\log_{10} E_{min}$ &  $\log_{10} E_{max}$ & IC-79 & IC-59 & IC-40 & IC-79 & IC-59 & IC-40 \\
\midrule[\heavyrulewidth]
\endfirsthead
\multicolumn{8}{c}%
{\tablename\ \thetable\ -- \textit{Continued from previous page}} \\[{0.5cm}]
\toprule
\multicolumn{2}{c}{ } & \multicolumn{3}{|c|}{North ($ 0^{\circ} < \delta \le 90^{\circ}$)}  & \multicolumn{3}{c}{ South ($ -90^{\circ} \le \delta \le 0^{\circ}$)} \\
\midrule
$\log_{10} E_{min}$ &  $\log_{10} E_{max}$ & IC-79 & IC-59 & IC-40 & IC-79 & IC-59 & IC-40 \\
\midrule[\heavyrulewidth]
\endhead
\bottomrule \multicolumn{8}{r}{\textit{Continued on next page}} \\[{0.5cm}]
\endfoot
\hline
\multicolumn{8}{l}{}\\[{0.1cm}]
\multicolumn{8}{l}{\parbox{\textwidth}{\textit{ Note. -- }    Solid-angle-averaged neutrino effective area for $\nu_\mu + \bar{\nu}_\mu$ in the north and south skies.  The first two columns indicates the limits of the energy bin so that $\log_{10} [E_{min}] < \log_{10} [E_{\nu} ({\rm GeV})]  \le \log_{10} [E_{max}]$.  The muon neutrino effective area is shown in units of m$^{2}$ for each of the three configurations.}}\\[{1cm}]
\endlastfoot

3.00 & 3.25 & 0.41 & 0.28 & 0.12 & 0.02 & 0.01 & 0.00 \\
3.25 & 3.50 & 1.11 & 0.78 & 0.34 & 0.06 & 0.04 & 0.01 \\
3.50 & 3.75 & 2.65 & 2.01 & 1.07 & 0.17 & 0.11 & 0.04 \\
3.75 & 4.00 & 5.87 & 4.65 & 2.56 & 0.46 & 0.32 & 0.11 \\
4.00 & 4.25 & 11.83 & 9.58 & 5.52 & 1.05 & 0.78 & 0.34 \\
4.25 & 4.50 & 21.77 & 18.36 & 12.00 & 2.52 & 1.83 & 0.74 \\
4.50 & 4.75 & 36.99 & 31.67 & 22.86 & 5.36 & 3.74 & 1.83 \\
4.75 & 5.00 & 58.47 & 50.86 & 34.85 & 10.88 & 7.99 & 3.26 \\
5.00 & 5.25 & 87.14 & 76.92 & 55.80 & 21.77 & 15.87 & 7.87 \\
5.25 & 5.50 & 121.76 & 108.50 & 81.50 & 42.85 & 30.24 & 15.34 \\
5.50 & 5.75 & 160.62 & 144.95 & 110.00 & 80.52 & 57.23 & 27.82 \\
5.75 & 6.00 & 205.52 & 187.38 & 141.89 & 147.57 & 106.95 & 53.59 \\
6.00 & 6.25 & 251.32 & 228.80 & 181.35 & 237.05 & 175.87 & 95.41 \\
6.25 & 6.50 & 300.92 & 280.14 & 216.07 & 360.95 & 275.64 & 172.67 \\
6.50 & 6.75 & 349.98 & 335.04 & 270.26 & 511.18 & 402.76 & 251.98 \\
6.75 & 7.00 & 406.74 & 379.00 & 298.75 & 701.98 & 549.56 & 366.78 \\
7.00 & 7.25 & 452.88 & 440.70 & 358.44 & 949.45 & 759.58 & 498.23 \\
7.25 & 7.50 & 497.98 & 481.35 & 419.92 & 1248.55 & 999.85 & 649.27 \\
7.50 & 7.75 & 561.75 & 531.64 & 482.86 & 1623.10 & 1324.28 & 834.44 \\
7.75 & 8.00 & 603.41 & 596.59 & 488.16 & 2084.37 & 1709.82 & 993.06 \\
8.00 & 8.25 & 660.84 & 660.13 & 535.53 & 2642.73 & 2164.94 & 1297.21 \\
8.25 & 8.50 & 719.94 & 732.64 & 520.84 & 3353.95 & 2779.39 & 1453.31 \\
\newpage
8.50 & 8.75 & 774.93 & 780.96 & 648.86 & 4227.17 & 3443.88 & 1608.56 \\
8.75 & 9.00 & 813.21 & 839.18 & 632.89 & 5307.43 & 4261.46 & 1746.93 \\
\end{longtable}
\end{center}

\section*{Acknowledgements}

We acknowledge the support from the following agencies: U.S. National Science Foundation-Office of Polar Programs, U.S. National Science Foundation-Physics Division, University of Wisconsin Alumni Research Foundation, the Grid Laboratory Of Wisconsin (GLOW) grid infrastructure at the University of Wisconsin - Madison, the Open Science Grid (OSG) grid infrastructure; U.S. Department of Energy, and National Energy Research Scientific Computing Center, the Louisiana Optical Network Initiative (LONI) grid computing resources; Natural Sciences and Engineering Research Council of Canada, WestGrid and Compute/Calcul Canada; Swedish Research Council, Swedish Polar Research Secretariat, Swedish National Infrastructure for Computing (SNIC), and Knut and Alice Wallenberg Foundation, Sweden; German Ministry for Education and Research (BMBF), Deutsche Forschungsgemeinschaft (DFG), Helmholtz Alliance for Astroparticle Physics (HAP), Research Department of Plasmas with Complex Interactions (Bochum), Germany; Fund for Scientific Research (FNRS-FWO), FWO Odysseus programme, Flanders Institute to encourage scientific and technological research in industry (IWT), Belgian Federal Science Policy Office (Belspo); University of Oxford, United Kingdom; Marsden Fund, New Zealand; Australian Research Council; Japan Society for Promotion of Science (JSPS); the Swiss National Science Foundation (SNSF), Switzerland; National Research Foundation of Korea (NRF).

\bibliographystyle{model1a-num-names}
\bibliography{<your-bib-database>}


\end{document}